\documentclass[a4paper,11pt]{article}
\pdfoutput=1
\usepackage{jinstpub} 
\usepackage{comment}
\usepackage{caption}
\usepackage{subcaption}
\usepackage{float}
\usepackage{url}
\usepackage{threeparttable}
\usepackage{amsmath}
\usepackage{hyperref}
\title{A quantitative assessment of Imaging High-Z and Medium-Z materials using Muon Scattering Tomography}
\author[a,1]{A.F.~Alrheli,\note{Corresponding author.}}
\author[a]{D.~Barker,}
\author[b]{C.~De~Sio,}
\author[b]{D.~Kiko\l{}a,}
\author[c]{A.K.~Kopp,}
\author[b,c]{M.~Mhaidra,}
\author[a]{J.P.~Stowell,}
\author[a]{L.F.~Thompson,}
\author[c]{J.J.~Velthuis,}
\author[a]{M.J.~Weekes.}

\affiliation[a]{University of Sheffield, Department of Physics and Astronomy, Hounsfield Road, Sheffield, S3 7RH, UK.}
\affiliation[b]{Warsaw University of Technology, Pl. Politechniki 1, 00-661 Warsaw, Poland}
\affiliation[c]{University of Bristol, School of Physics, HH Wills Physics Laboratory,
Tyndall Avenue, Bristol BS8 1TL, UK.}

\emailAdd{afalrheli1@sheffield.ac.uk}

\abstract{
Muon Scattering Tomography (MST) has been shown to be a powerful technique for the non-invasive imaging of high-shielded objects. We present here the application of the MST technique to investigate two types of nuclear waste packages, a small-steel drum and a large nuclear waste cask, namely, a CASTOR V/52. We have developed a quantitative method using the contrast-to-noise ratio (CNR) to evaluate the performance of an MST detector system in differentiating between high-, medium-, and low-Z materials inside nuclear waste packages with different shielding types. This study reveals that our MST detector system is able to differentiate between a (10 $\times$ 10 $\times$ 10 cm$^3$) uranium cube, embedded within a concrete matrix inside the small-steel drum,
and regions of background signal in six hours of muon exposure time with a CNR value of 3.1$\pm$0.2. During our investigation of the highly-shielded cask, the reconstructed images of the cask contents indicated the ability of our system to detect irregular baskets, such as empty baskets, with a CNR value of 5.0$\pm$0.3 after 30 days of muon exposure. These studies were done using a Monte Carlo simulation tuned to the performance of resistive plate chambers (RPCs) based muon tomography system built by the University of Bristol, which had a reported position resolution of 350 micron.

Here we also report the dependence of the performance on the position resolution. We argue that using a combination of RPC and drift chambers (DC) detectors with 700 micron and 4 mm position resolutions respectively is able to generate tomographic images of well-shielded materials in a few hours of muon exposure time. With these position resolutions, our system needs six hours of muon exposure time to produce a good quality image of a cube of uranium with side-length of 10 cm shielded by a concrete matrix with CNR value of 2.4$\pm$0.25.  
}


\begin{document}
\maketitle
\flushbottom
\section{Introduction} \label{sec:intro}
 Investigating well-shielded objects ideally requires a non-destructive method, especially when dealing with hazardous materials such as radioactive materials inside nuclear waste packages. Using a non-invasive technique to assay nuclear waste packages could beneficially reduce the cost involved in opening the investigated packages, as well as mitigate against any potential risk of being exposed to ionising radiation. Muon scattering tomography (MST) is a non-destructive imaging method that has proven valuable for imaging hidden objects in many applications, such as nuclear safeguards and nuclear waste management \cite{Morris2008a,Frazao2016}. 
 
 The efficacy of employing MST as an imaging method is ascribed to several factors, e.g., cosmic muons are a naturally-occurring source, and, furthermore, the penetration level of cosmic muons is much higher than that of other conventional imaging methods, such as X-rays.
 
Globally, various scientific groups have developed detectors to exploit the MST technique to reconstruct a 3D image of potentially hazardous materials \cite{Borozdin2003,Pesente2009}. Recently, several studies have demonstrated the effective performance of the MST method in identifying high-Z materials surrounded by shielding materials, such as uranium cubes embedded inside a cemented nuclear waste drum \cite{Frazao2019,Weekes2021}.

However, the performance of the MST technique and associated algorithms in reconstructing a 3D image of the volume of interest can be adversely affected by several factors, such as a thicker and/or higher-density shielding matrix. Detailed simulation studies are required to extract information about the roles of each variable that is likely to enhance or degrade the performance of the MST technique. 
In the past, conventional statistical methods have been applied to evaluate the methods that are used; for example, the automated Receiver Operating Characterisation (ROC) curve is traditionally used to analyse the performance of an automated binary classifier in correctly identifying a hazard. 
This automated method can be affected by irregularities in the shielding matrix,
leading to an incorrect report of the presence of a potential hazard (false positive). 

In this paper, several Monto Carlo (MC) simulation studies were performed to assess the MST technique and algorithms used to reconstruct images of various materials. It is desirable to develop a quantitative method to evaluate the efficacy of the feature resolution outputs when imaging different materials with different atomic numbers inside different nuclear waste packages. A simple statistical contrast-to-noise ratio (CNR) method \cite{Prince2006} is used to evaluate the capability of the system and the chosen algorithms in differentiating between two regions inside the volume of interest. Specifically, regions inside a small cemented-matrix drum representing: a region of hazardous high-Z material, a region of non-hazardous high-Z material, a region of non-hazardous medium-Z material, and a region of non-hazardous low-Z material are considered. 

To understand whether the shielding type affects the performance of the chosen algorithm, this test was also used to compare several regions inside a simulated ductile-iron shielded large-scale cask. The considered regions inside the large-scale cask represent: a region of fully-loaded basket with the fuel assemblies, a region of half-loaded basket with the fuel assemblies, a region of fully-loaded basket with non-hazardous materials and a region of an empty basket.

\section{The Muon Scattering Tomography Method} \label{sec2}
The principle of MST as an imaging technique is briefly described in this section. Muons are charged particles that originate as a result of interactions of primary cosmic rays with the Earth’s upper atmosphere.  The significant flux of cosmic muons, namely 10,000 muons $\rm{m^{-2} min^{-1}}$ at sea level makes it possible to use them in imaging technologies \cite{Beringer2012}. Despite the fact that a muon has a large mass, it can undergo multiple scattering events as it travels through material. In principle, the scattering angle of muons is inversely proportional to the radiation length of the material that the muon traverses (X$_\mathrm{o}$). Statistically however, the distribution of the projected scattering angle of muon through a material with a thickness of X is approximately gaussian, with a width $\sigma$ given by:
\begin{equation}
    \sigma \approx \frac{13.6 \mathrm{MeV}}{\beta c p} \sqrt{{X}/{X_o}}
\end{equation}
where $\beta$c is the muon velocity, p is the muon momentum \cite{Lynch1991} , and X$_\mathrm{o}$ is determined by:

\begin{equation}
    X_o = \frac{716.4 \mathrm{A}}{\rho  \mathrm{Z(Z+1)} \ln(287/\mathrm{\sqrt{Z})}} 
\end{equation}
where $\rho$ represents the material density, Z is the atomic number of the material and A is the atomic mass number \cite{Eidelman2004}.

\begin{figure}[H]  
\begin{center}
      \includegraphics[scale=0.25]{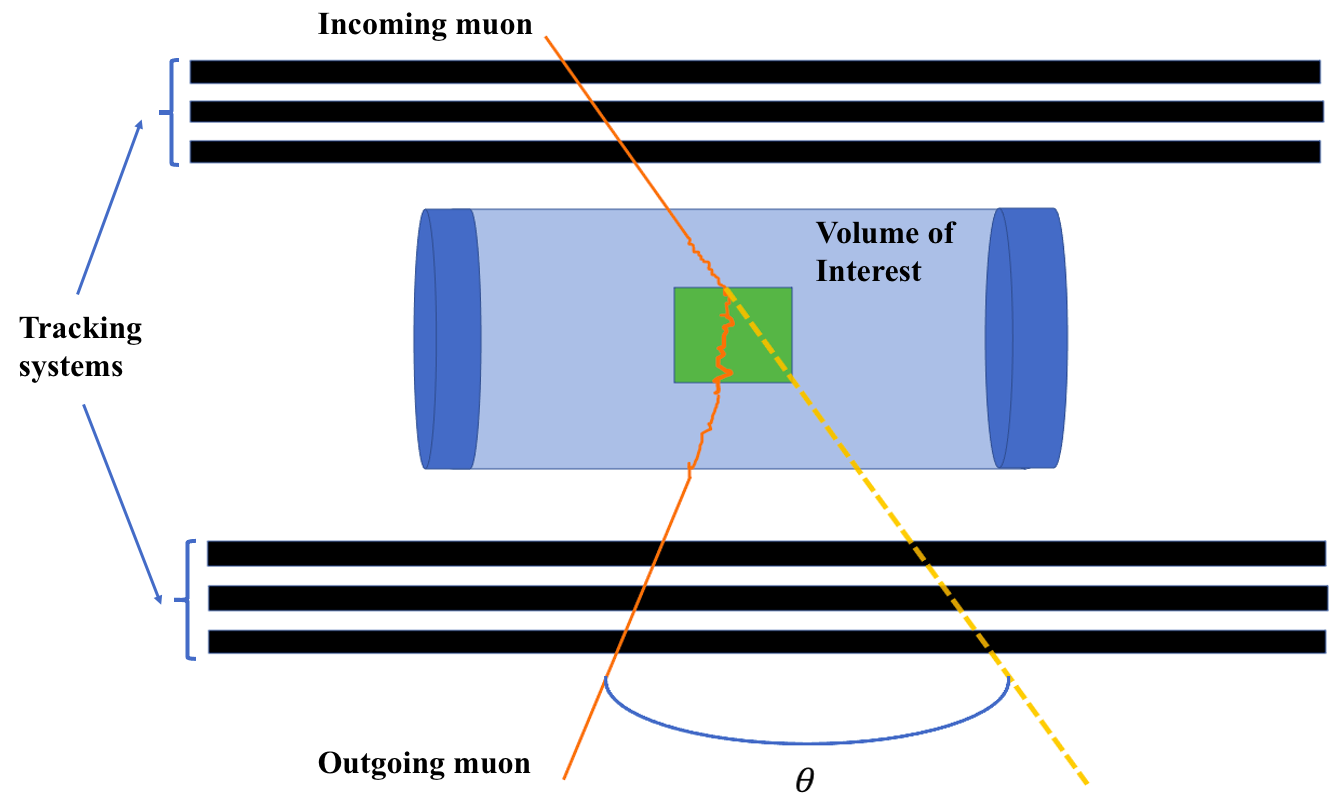}
     \caption{Schematic depicting a generic muon scattering tomography system comprising tracking chambers placed above and below a volume of interest containing a cube of high-Z material in green. The solid orange lines represent the trajectories of the incoming and outgoing muons, hence the scattering angle ($\theta$) can be determined.} 
     \label{figure1_label}
 \end{center}    
\end{figure}

Unlike other imaging techniques, such as computed tomography (CT), 
which employ a focussed narrow beam aimed at a target,
MST simply places the volume of interest between two tracking systems to track the muon trajectory as it enters and exits the object (see Figure \ref{figure1_label}). In order to produce a 3D density map of the investigated volume, both tracking systems should consist of multiple layers to provide muon hits in the X-Z and Y-Z planes. Reconstruction of the muon trajectories for muons both entering and exiting a volume of interest can be obtained; hence, a 3D density map of the volume can be produced by reconstructing the muon scattering angle distributions.

The muon trajectories are reconstructed using dedicated algorithms. In this paper, we consider three frequently-used algorithms; 
the simple Point of Closest Approach (PoCA) algorithm \cite{Riggi2013} (see section \ref{subsection3.1}), the Angle Statistics Reconstruction (ASR) algorithm \cite{Stapleton2014} (see section \ref{subsection3.2}), and the Binned Clustering (BC) algorithm \cite{C.Thomay2013} (see section \ref{subsection3.3}). All these algorithms divide the volume of interest into a 3D cubic-voxel grid with a side length of typically 10 mm for each voxel. A discriminator score for each voxel is then extracted from all the muon trajectories that have traversed through the volume of interest. Subsequently, all regions containing high-Z, medium-Z and low-Z material inside the volume of interest can be identified.

\section{The Reconstruction Algorithms} \label{section3}
\begin{figure}[htb]
\begin{subfigure}{.5\textwidth}
  \centering
  \includegraphics[width=0.9\linewidth]{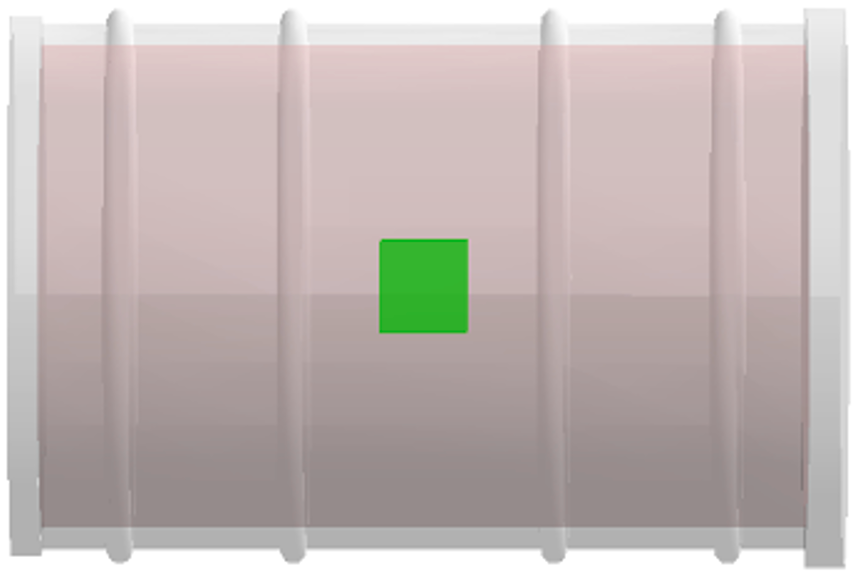}  \caption{}
  \label{fig:sub-first}
\end{subfigure}
\begin{subfigure}{.5\textwidth}
  \centering
  \includegraphics[width=1.0\linewidth]{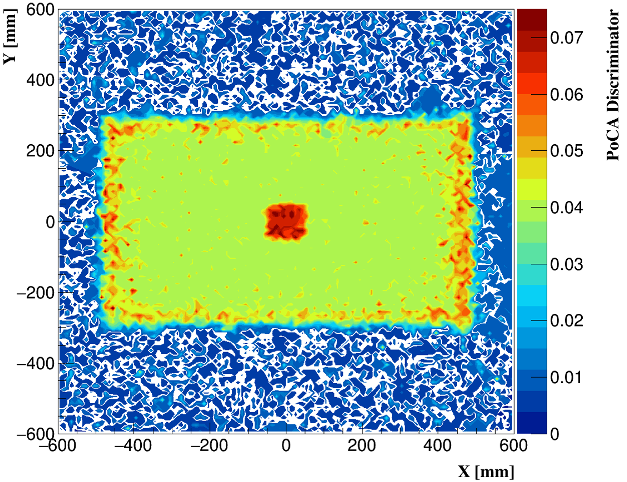}  
  \caption{}
  \label{fig:sub-second}
\end{subfigure}

\begin{subfigure}{.5\textwidth}
  \centering
  \includegraphics[width=1.0\linewidth]{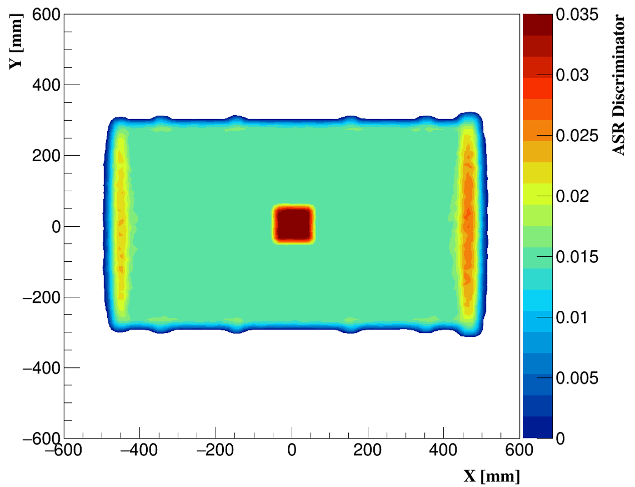}  
  \caption{}
  \label{fig:sub-third}
\end{subfigure}
\begin{subfigure}{.5\textwidth}
  \centering
  \includegraphics[width=1.0\linewidth]{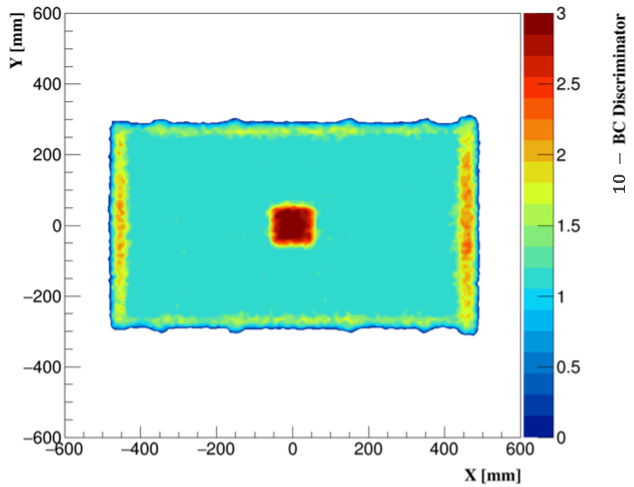}  
  \caption{}
  \label{fig:sub-fourth}
\end{subfigure}
\caption{(a) a 10 cm side length tungsten cube placed in the centre of a simulated cement matrix nuclear waste drum. The 2D projected output of the same cube inside the small drum using the (b) PoCA (c) ASR and (d) BC algorithms. The exposure time was 25 days equivalent.}
\label{figure2_label}
\end{figure}

This section will explain in detail the methods used by the three reconstruction algorithms considered in this work. In addition, an example of imaging a tungsten cube with a side length of 10 cm embedded inside a small nuclear waste drum will be used to compare the performance of each algorithm (see Figure \ref{figure2_label}(a)). In order to visualise the tungsten cube inside the drum, the 3D map produced by each algorithm is sliced into a 2D projection.
\subsection{Point of Closet Approach (PoCA) Algorithm} \label{subsection3.1}
As an algorithmic reconstruction imaging method, the Point of Closest Approach (PoCA) algorithm is a straightforward method that offers less complexity in computational analysis, which makes it commonly used in MST techniques. It assumes that when a muon travels through the 3D voxel grid, it undergoes multiple scatterings, which are then approximated to a single scattering at a single position (vertex). This single scattering position is located by extrapolating the entrance and exit points of muon tracks through the volume of interest. The voxel containing the minimum distance between the tracks is defined as the single scattering voxel.

In theory, the muon is likely to have a large scattering angle when it encounters high-Z material inside the investigated volume. Inside the 3D cubic voxel grid, a value is attributed to each voxel which is determined by weighting the average angle for all muon trajectories whose point of closest approach is located in that voxel, this value will be referred to in this paper as the PoCA discriminator. Hence, materials with a high Z can be distinguished from other materials with a lower Z. However, the PoCA method is susceptible to adding noise as a result of the assumption that a single large scatter has taken place (see Figure \ref{figure2_label}(b)). This assumption has a number of weaknesses, e.g., some muon tracks might be incorrectly extrapolated and some PoCA points may occur outside the volume of interest. Such instances of muon track mis-reconstruction within the volume of interest can lead to inaccurate information which may be interpreted as an extreme scattering vertex inside a voxel. 

\subsection{Angle Statistics Reconstruction (ASR) Algorithm} \label{subsection3.2}
Unlike the PoCA method, the ASR algoritm assumes that when a muon travels through the 3D voxel grid, it is likely to experience many small scatters. The ASR thus avoids the underlying assumption of the PoCA algorithm that a muon only scatters inside a single-voxel. The Angle Statistics Reconstruction (ASR) algorithm was developed to mitigate the effects of using the PoCA method’s inaccurate approximation of the muon trajectories.
This has been achieved by applying a minimum chosen distance ($D_r$) between the reconstructed muon trajectories and the centre (c) of a voxel, therefore only voxels that lie within the chosen distance are considered. Any voxel that is located beyond the chosen distance  $D_r$ will be neglected. The minimum distance is determined by:
\begin{equation}
    D_r = \max ( \min ||a(z) - c||, \min||b(z) - c||)
\end{equation}

where a(z) and b(z) represent the fitted trajectories of the incoming and the outgoing muons, respectively.

A threshold distance of ($d_{th}$) is chosen, ideally it is the same size as a voxel so that all voxels that have $D_r$<$d_{th}$ will be assigned a discriminator score.
For each voxel and each muon with momentum of (P), the projected scattering angles on the x-axis and y-axis ($\theta_x$ and $\theta_y$ respectively) are used to generate two scores $S_1$=(| $\theta_x$|~$\cdot$~$\widetilde{p}$) and $S_2$=(| $\theta_y$|~$\cdot$~$\widetilde{p}$), where $\widetilde{p}$ is the muon's momentum according to $\widetilde{p}$ = $\frac{P}{P_{norm}}$ , and ${P_{norm}}$ = 3 GeV . This is repeated for all muons passing through the object of interest resulting in a distribution of the $S_1$ and $S_2$ scores for each voxel. Scores are only added to a voxel's distribution if the entering/exiting muon trajectories pass within $d_{th}$ of the voxel.  
For each voxel the final distribution of scores is taken and an ASR discriminator score is assigned to be that of the third quartile (0.75) of the distribution and this value will be referred to in this paper as the ASR discriminator. These final discriminator scores are subsequently used to locate voxels in which high-Z materials might be present. (Figure \ref{figure2_label}(c)) demonstrates how, by excluding outlier events with extreme scattering, the ASR algorithm has successfully reduced noise resulting in a clearer image. 

\subsection{Binned Clustering (BC) Algorithm} \label{subsection3.3}
The BC algorithm \cite{C.Thomay2013} is based on the fact that the density of high angle scattering vertices is higher in high-Z materials. After assigning the scattering vertices to a voxel, the scattering angles are ordered and only the N most highly scattered vertices are considered to calculate the metric distance between two vertices weighted by their scattering angle. Voxels containing a number of vertices less than N are neglected. The choice number of N is important as it can effect on the reconstructed image. For instance, higher value of N might cause a distortion of the reconstructed image as a result of removing more voxels containing lower number of vertices than the chosen N.

\begin{figure}[hbt]
\begin{center}
      \includegraphics[scale=0.5]{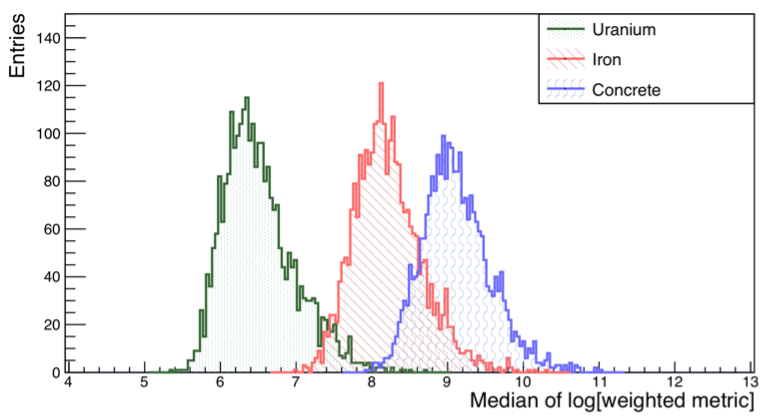}
     \caption{Comparison of distributions of the BC algorithm discriminator, for 15 cm cubes of uranium, iron and concrete. Higher discriminator values correspond to low Z material.} 
     \label{figure2s_label}
 \end{center} 
\end{figure}

For each pair of vertices V$_i$ and V$_j$ within the investigated volume, the weighted metric distance is determined by:

\begin{equation}
    {\widetilde{m}_{ij}} = \frac{||V_i - V_j||}{(\theta_{i} \widetilde{p_i}) \cdot (\theta_{j} \widetilde{p}_{j})}
\end{equation}
where, $\theta_i$ , $\theta_j$ are the scattering angles for muons i and j respectively in vertex $V_i$ and $V_j$. For muon i,  $p_i$ is the muon momentum and $\widetilde{p_i}$ is the muon momentum according to $\widetilde{p_i}$ = $\frac{P_i}{P_{norm}}$ where $P_{norm}=3$ GeV/c.
The median of the distribution of log(${\widetilde{m}_{ij}}$) inside a voxel is used as the discriminator value for that voxel and this value will be referred to in this paper as the BC discriminator. This is expected to be lower if the target is a high-Z material as the average distance between high scattering vertices is smaller and the scattering angles higher in high-Z materials. Conversely, if the target is a low-Z material, the median would be higher (see Figure \ref{figure2s_label}). The BC discriminator output for imaging the tungsten cube inside the waste drum reflects good contrast when compared to the background (see Figure \ref{figure2_label}(d)).
\section{Monto Carlo Simulations} \label{section4}
\subsection{The MST detector system} \label{subsection4.1}
The detector system was simulated using a cosmic ray simulation platform (CRESTA) \cite{Steer} built on the GEANT4 high-energy particle physics simulation toolkit \cite{Agostinelli2003}. The system consists of two identical detector systems placed above and below the volume of interest with a gap of 105~cm when investigating the small cemented-matrix drum. The two tracking systems are placed with a gap of 580~cm apart when investigating the large CASTOR cask. The MST detector system combines scintillating triggers and two types of gaseous particle detectors.

\begin{figure}[H]  
\begin{center}
      \includegraphics[scale=0.3]{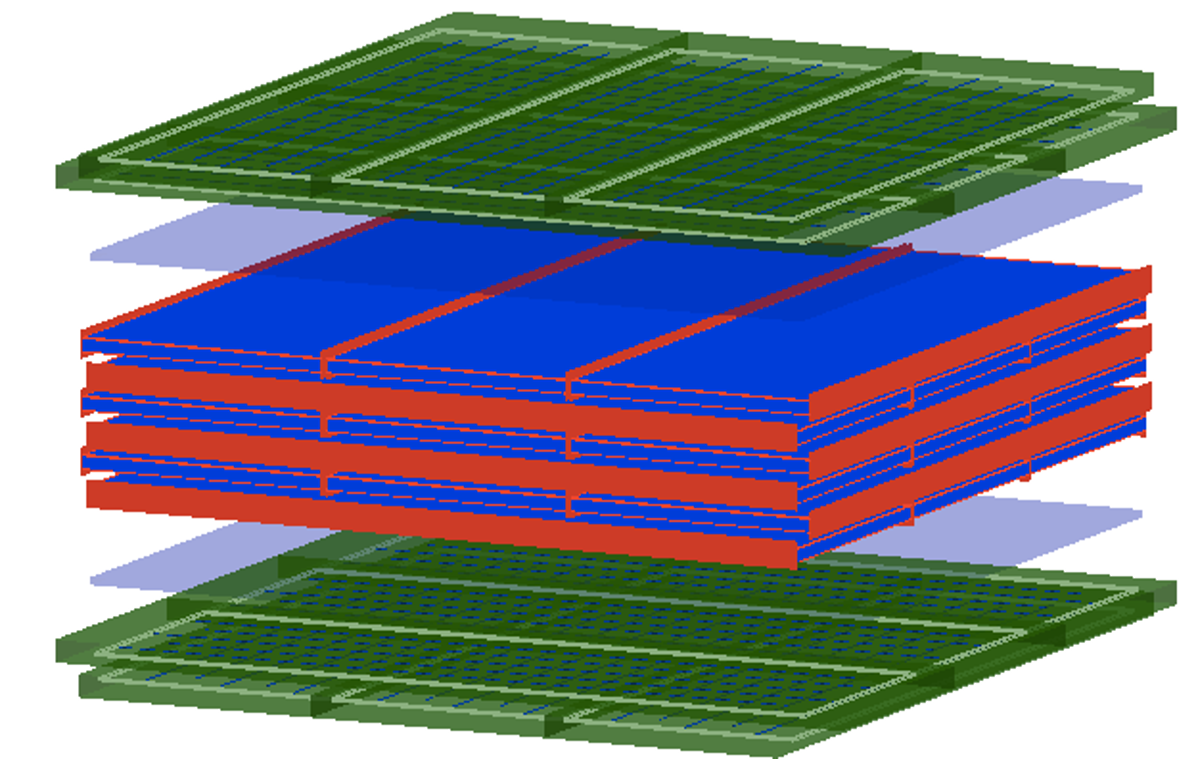}
     \caption{Diagram the simulated tracking system with active area of approximately 2~m x 2~m, indicating 4 layers of RPCs (in green) and 6 layers of DCs (in blue and red). The light blue panels represent the scintillator triggering system.} 
     \label{figure3_label}
 \end{center}    
\end{figure}

Each tracking system is arranged in 10 layers: six layers of Drift Chambers (DC) and four layers of Resistive Plate Chambers (RPC) with spatial resolutions of $\sim$2 mm and $\sim$0.35 mm, respectively (see Figure \ref{figure3_label}). The 10 layers detectors are arranged to provide muon trajectories in the x and y planes: five layers placed in the X-Z plane and five layers are rotated by 90 degrees to provide muon hits in the Y-Z plane. Muons are generated by the CRY cosmic ray shower generator library \cite{Hagmann2007}. The detectors record muon hit position information, while the scintillator detectors provide a trigger of when a muon has passed through the system. This MST simulated system will be used to test the performance of algorithms described in section \ref{section3} by reconstruction of a variety of materials which have different atomic numbers embedded inside two different types of nuclear waste packages with different shielding types and sizes.

\subsection{Simulation of Nuclear Waste Packages} \label{subsection4.3}
The algorithmic methods detailed in section \ref{section3} are applied to the MC simulations to investigate their performance 
in distinguishing high-Z materials stored inside different shielding matrices. Two nuclear waste packages were considered in this study: a small cemented-matrix nuclear waste drum and a large ductile-iron shielded V/52 CASTOR cask.

The small waste drum simulates the disposal of a mixture of radioactive wastes surrounded by a concrete matrix. The drum is made of steel ($\sim$92$\%$ iron and 2$\%$ carbon) of 96~cm length and 57.4~cm diameter. It is filled with concrete to a total diameter of 52.4~cm and density 2.3~g/cm$^3$ (see Figure \ref{figure2_label} (a)).

\begin{figure}[htb]
\begin{subfigure}{.6\textwidth}
  \centering
  \includegraphics[width=0.7\linewidth]{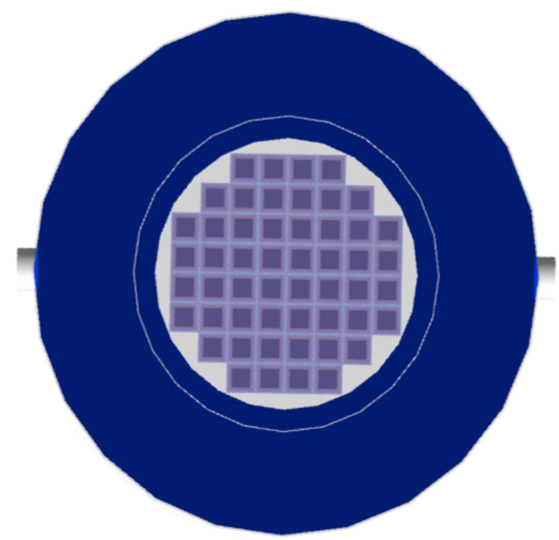}  \caption{}
  \label{fig:sub-first3}
\end{subfigure}
\begin{subfigure}{.4\textwidth}
  \centering
  \includegraphics[width=0.6\linewidth]{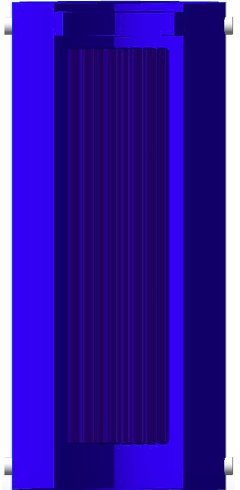}  
  \caption{}
  \label{fig:sub-second3}
\end{subfigure}

\begin{subfigure}{.5\textwidth}
  \centering
  \end{subfigure}
  \caption{(a) top and (b) side views of the simulated V/52 CASTOR cask accommodating the 52 waste baskets. The lid and the base removed for visualisation purposes.}
  \label{figure4_label}
\end{figure}

To understand the effects of shielding thickness on the algorithmic outputs, a large V/52 CASTOR cask with denser shielding type has also been simulated (see Figure \ref{figure4_label}). The cylinder-shaped V/52 cask is made of ductile-iron ($\sim$94$\%$ iron, 0.033$\%$ carbon, 0.004$\%$ copper) 
with a height of 5.54~m and a total diameter of 2.44~m. A cavity of 1.42~m diameter and 4.55~m height inside the centre of the cask is designed to accommodate the baskets for the fuel assemblies, which are surrounded by nearly 1 m of ductile-iron shielding. The cavity is designed to store 52 baskets that accommodate UO$_2$ ( $\sim$ 88.2 $\%$ uranium and 11.8 oxygen) fuel assemblies that originate from Boiling Water Reactors (BWR). The simulated box-shaped baskets have a length of 4.48~m and are arranged across a grid of eight columns and eight rows. A pair of trunnions is also simulated at the top and the end bottom of the CASTOR. These trunnions are bolted and only be used for the attachment of handling equipment.  
\newline
\newpage
\section{Performance Tests} \label{section5}
We developed a method to quantitatively compare the outputs of all of the algorithms to each other.  The contrast to noise ratio (CNR) method is applied to compare two regions in the reconstructed image of the investigated drum, such as a region containing high-Z material against another region containing a background signal.
\subsection{Contrast-to-Noise Ratio (CNR)} \label{subsection5.1}
To assess the reconstructed image quality, the CNR method was developed to evaluate the capability of the considered algorithms in differentiating between low-contrast, medium-contrast, and high-contrast regions inside the investigated volume.

A high value CNR indicates the algorithm is able to distinguish between the two regions under comparison. Likewise, a low CNR value between the compared regions, is the result of the algorithm's inability to distinguish between the two regions. The CNR value for two regions, A and B, is calculated to understand the feature resolution of the MST system and algorithm under consideration, and is given by:

\begin{equation}
    \mathrm{CNR} = \frac{|\mu_{A} - \mu_{B}|}{\sqrt{\sigma_{A}^{2} + \sigma_{B}^{2}}}
\end{equation}
where $\mu_A$ is the mean of region A’s signal and $\mu_B$ is the mean of region B’s signal. $\sigma_{A}$ and $\sigma_{B}$ are the standard deviations of the signals in region A and region B, respectively.

\begin{figure}[H]  
\begin{center}
      \includegraphics[scale=0.36]{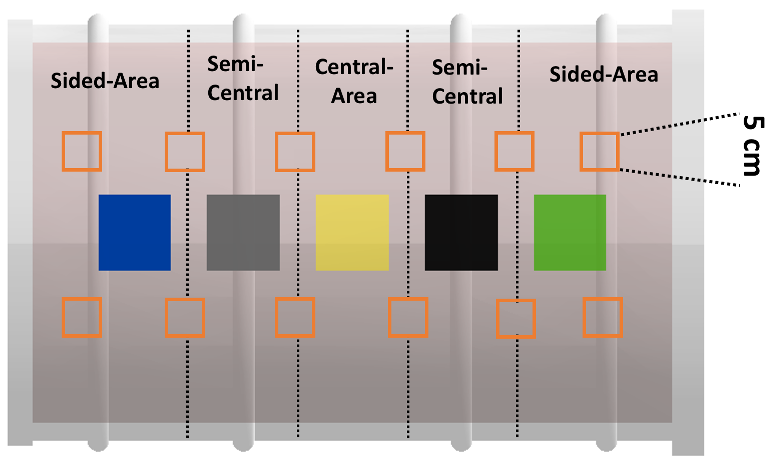}
     \caption{The five cubic materials with a side length of 10 cm placed within concrete inside a steel drum. From the left, Aluminium, Iron, Copper, Lead and Uranium. The small orange squares represent the chosen background regions from both the side and central areas of the drum.} 
     \label{figure5_label}
 \end{center}    
\end{figure}

Initially, an array of five cuboid materials (see Table \ref{table:1}) classified as either hazardous or non-hazardous are simulated each with a side length of 10~cm and embedded inside the small cemented matrix waste drum along the x-axis (see Figure \ref{figure5_label}). The target materials were chosen to vary in terms of atomic number and density.

\begin{table}[htp]
\centering
\begin{tabular}{ |c|c|c|  }

  \hline
   \textbf{Target Material}& \textbf{Atomic Number (Z)} & \textbf{Density g/cm$^3$}\\
   \hline
   Uranium   & 92    &18.95\\ 
   \hline
   Lead&   82  & 11.35\\
   \hline
    Copper &29 & 8.96\\
    \hline
    Iron    &26 & 7.87\\
    \hline
    Aluminium &11   &2.699\\
    \hline
\end{tabular}

\caption {Characteristics of the target materials under consideration.}
\label{table:1}
\end{table}
For each algorithm described in section \ref{section3}, the 3D density map of the drum contents is sliced into a 2D projection, and then the regions inside the drum are located based on the discriminator score for each method. Comparing each of the five regions to the background region should reveal the ability of each algorithm to locate high-density, medium-density and low-density materials inside the drum. For instance, the CNR value between the copper region and the background region allowed us to test the ability of each algorithm used to distinguish the copper cube within the background regions. Then, a hazardous high-Z material region (uranium) is compared to the non-hazardous copper and iron regions, which are classified as medium-Z materials. Finally, comparing two regions that contain materials with almost similar atomic numbers and densities (copper and iron) would be helpful in understanding the efficacy of each algorithm.

The size and location of material within the drum may also affect the algorithm's performance. Hence, three simulations of the same materials were done with different sizes and locations of the target materials, namely cubes with side dimensions of 7~cm, 10~cm and 13~cm. The positions of the materials are classified as central, semi-central and side locations, as illustrated in Figure~\ref{figure5_label}.

In the second study, the 52 fuel assembly baskets in a CASTOR V/52 nuclear waste container were simulated to accommodate different materials, namely uranium oxide, lead, and copper, whilst one basket was intentionally left empty (see Table \ref{table:2}). The CNR test was subsequently extended to understand whether the performance of each algorithm would be affected by a thicker and more shielded cask.

\begin{table}[htb]
\centering
\begin{threeparttable}
\begin{tabular}{ |c|c|c|  }

  \hline
   \textbf{Basket Content}& \textbf{Number of Baskets} & \textbf{Density g/cm$^3$}\\
   \hline
   Uranium Oxide (Fully-loaded) & 48    &10.97\\ 
   \hline
   Uranium Oxide( Half-loaded)&   2  & 10.97\\
   \hline
    Lead &1 & 11.35\\
    \hline
    Copper &1 & 7.87\\
    \hline
    Empty &1  &0.0012\\
    \hline
\end{tabular}
\end{threeparttable}
\caption {Details of the target materials placed inside the simulated CASTOR V/52 waste drum.}
\label{table:2}
\end{table}

 The CNR test conveys information about the feature resolution of the algorithms in order to distinguish the contents of each basket individually and separate abnormal baskets (e.g., empty) from the UO$_2$ fully-filled baskets. The feature resolution can be tested by comparing the CNR value for a basket that accommodates UO$_2$ fuel assemblies with that from another basket filled with pellets of materials classified as a non-hazard, such as lead or copper. To test the size resolution, two baskets had up to $50\%$ of their normal capacity unloaded and positioned randomly throughout the CASTOR. Half of the fuel assembles are removed from different positions inside each half-loaded basket, including both the side and central; comparing two regions of the half-loaded basket to the fully loaded basket assesses the ability of the algorithm to detect any irregularities within the basket to be evaluated.
 
 
 There is a possibility of having different CNR values if two similar comparable regions are located in the same area, such as two UO$_2$ fully-filled baskets in the central area compared to an empty basket. Despite having similar contents in the same area as the full-loaded baskets, the comparison can be affected by the standard deviations of the signals in the two regions. To overcome this issue, for each region of interest (e.g., an empty basket), the average discriminator of the region of the empty basket will be compared to the average discriminators of the region of eight fully loaded baskets surrounding the empty one.
 \newline
 \begin{figure}[htb]  
\begin{center}
      \includegraphics[scale=0.38]{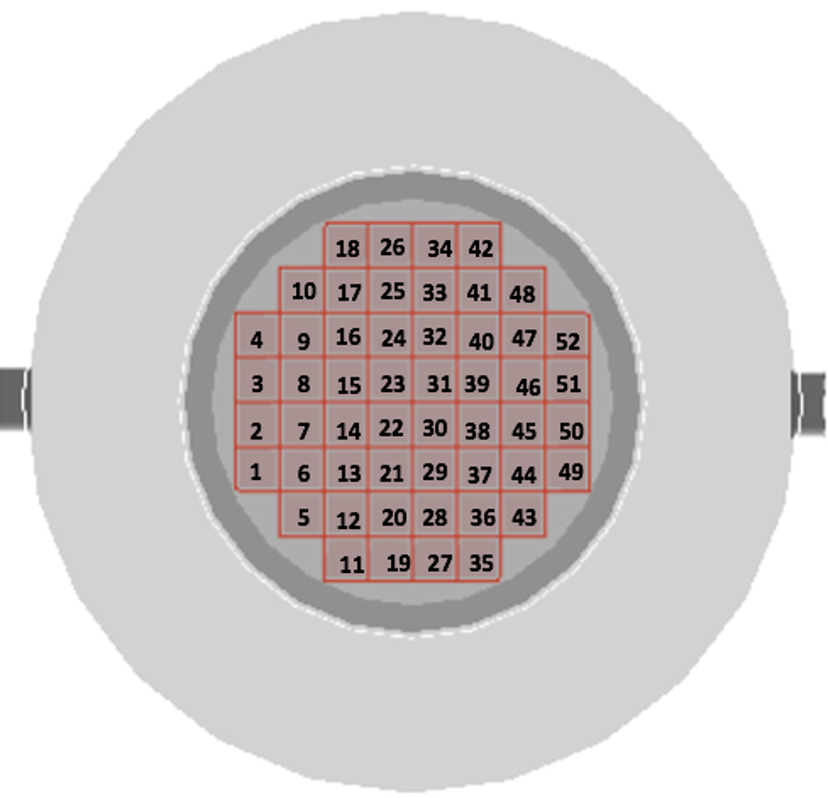}
     \caption{Top view of the V/52 CASTOR, showing all the baskets and labelling them with numbers from 1 to 52 (the lid and the base are removed for illustrative purposes).} 
     \label{figure6_label}
 \end{center}    
\end{figure}
 
 It is important to understand the effect that the location of the fuel assembly basket inside the cask has on the CNR values for each algorithm. Hence, the target (abnormal) basket will be classified depending on its location inside the cask as either a central-area or side-area basket.  Abnormal baskets are either filled with non-hazard materials, half-loaded UO$_2$ fuel assemblies, or are completely empty. For the side-area basket study a single abnormal basket was placed in the side area of the cask (basket no. 7) and then compared with the eight fully loaded fuel assembly baskets surrounding it for each of the fill scenarios detailed above (see Figure \ref{figure6_label}). Similar simulations for the same materials placed in a basket located in the central area (basket no. 30) of the cask 
 were also carried out for the central-area studies.  Finally, for each target basket, the final value of the CNR is obtained by calculating the average of the CNR values measured when the material is positioned in the side and central areas of the cask.
 
 \subsection{Minimum CNR (mCNR) value} \label{subsection5.2}
When comparing results from the imaging of different materials in both the small steel drum and CASTOR V/52 cask a consistent metric to compare results is required. The concept of a ``minimum CNR'' (mCNR) is thus introduced to distinguish between two regions. Good quality of the reconstructed images is highly desirable in this study, hence we have applied two conditions to determine the mCNR values for all methods to differentiate between region A and region B:
 \begin{itemize}
     \item The structures of regions A and B must be fully reconstructed ( method's discriminator $>$ 0).
     \item The mean discriminator of region A must be separated from that for region B by at least $\sigma_A + \sigma_B$ (see Equation \ref{mCNR}) where $\sigma_A$ and $\sigma_B$ are the errors on the discriminators for regions A and B respectively.
 \end{itemize}
 \begin{equation}
    \mu_A - \mu_B > \sigma_A + \sigma_B
    \label{mCNR}
 \end{equation}
 The mCNR value is defined by the minimum CNR value produced by each algorithm in which two comparable regions can be distinguished from each other regardless of the muon exposure time.
 In this test, a cube of medium-Z material (copper) with side-length of 10~cm is reconstructed and compared to equally-sized regions containing a background signal (henceforth ``multi-regions''). The mCNR values for the ASR and the BC algorithms are calculated by comparing the copper cube against the background region. The PoCA method was unable to locate the copper cube, hence the mCNR of the PoCA method is calculated by comparing a 10~cm side-length uranium cube against the background region.

 \begin{figure}[htp]
\centering
\includegraphics[width=.65\textwidth]{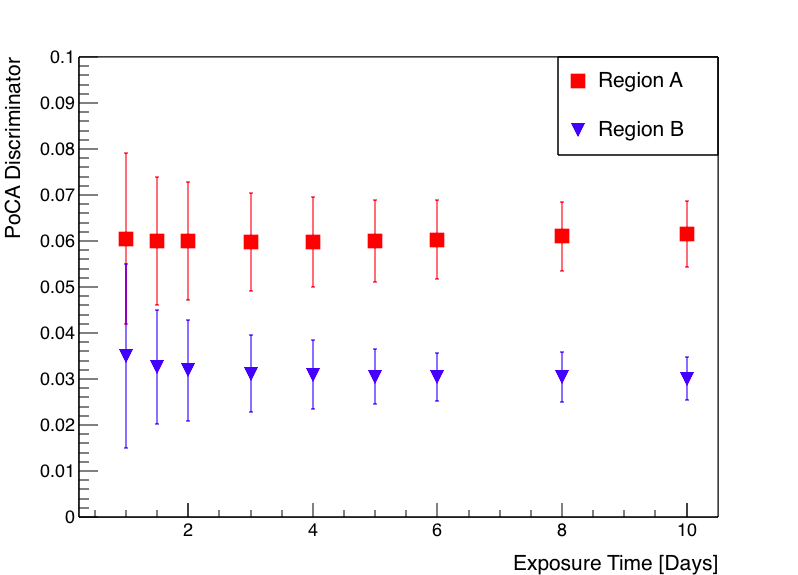}\quad
\medskip

\includegraphics[width=.485\textwidth]{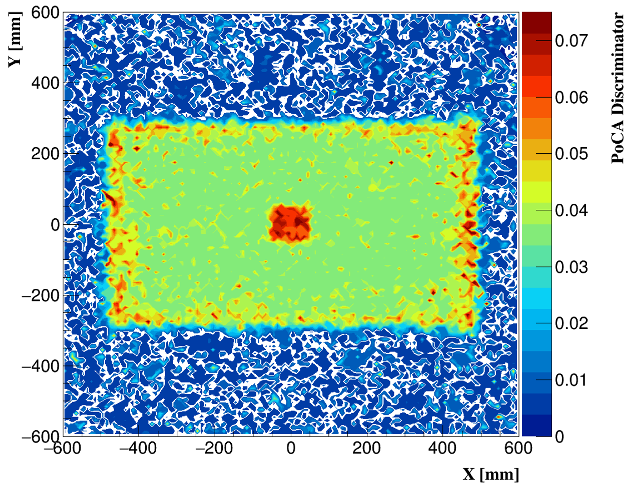}\quad
\includegraphics[width=.485\textwidth]{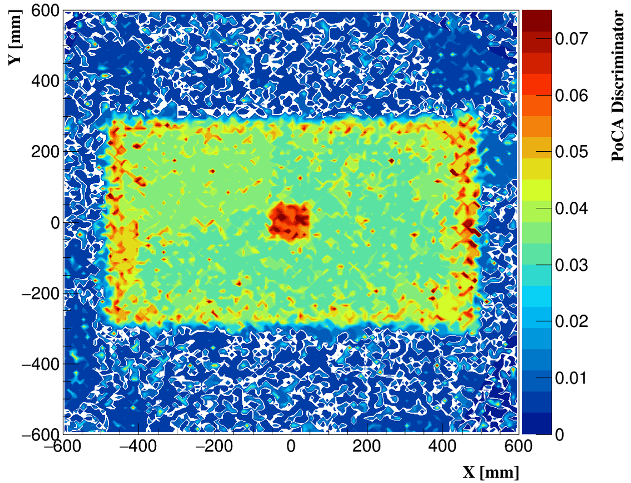}\quad
\includegraphics[width=.485\textwidth]{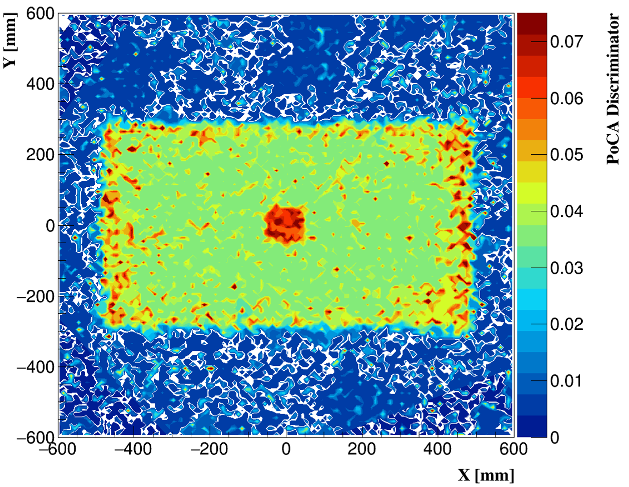}\quad
\includegraphics[width=.485\textwidth]{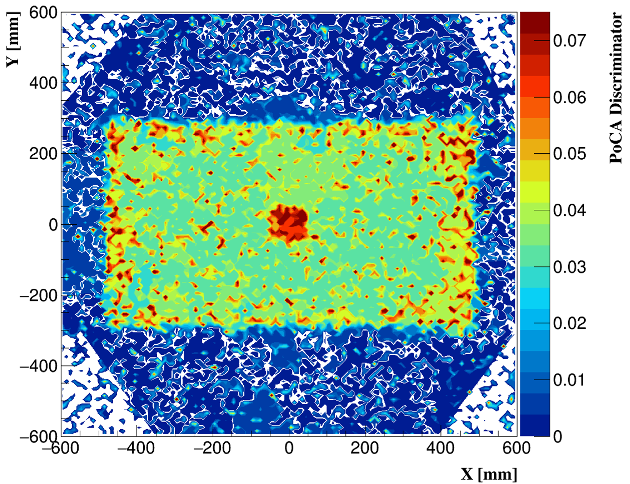}

\caption{The top figure shows the comparison between the PoCA average discriminators for a uranium cube with 10~cm side-length (region A) against a 10~cm side-length of multi-regions containing a background signal (region B). The middle and the bottom figures represent the outputs of the PoCA algorithm after 3 days (middle-left), 2 days (middle-right), 1.5 day (bottom-left) and 1 day (bottom-right) of muon exposure time.}
\label{figure7_label}
\end{figure}
The BC method can distinguish the copper cube from the background region in 5 days with a CNR value of 1.4, the BC discriminators for the two comparable regions are separated by more than the sum of their errors. Whereas, ASR only needs 18 hours to distinguish the copper cube from the background region with a CNR value of 1.4, the ASR discriminators for the two comparable regions are separated by more than the sum of their errors. This means that the ASR and BC algorithms need a CNR value $\geq$ 1.4 to be able to distinguish between the two regions.

 The PoCA algorithm can achieve the minimum distinguishable limit between the U cube region (region A) and the background region (region B) after 1.5 days of muon exposure time with a CNR value of 1.4 (see Figure \ref{figure7_label}). This means the PoCA method cannot distinguish between the two regions if the CNR value is lower than 1.4. For example, the output of PoCA after only 1 day of exposure time produced a low quality image with more variation in the background regions, and thus the CNR value between the comparable regions reduced to 1.0 which is insufficient to differentiate between the two regions (see Figure \ref{figure7_label}). 
 \newpage
 \section{Results and Discussion}
 \subsection{Application of the CNR test to a small nuclear waste drum}
The three reconstruction algorithms discussed earlier were each used to image five target materials that varied in density from 2.699~g/cm$^3$ (aluminium) to 18.59~g/cm$^3$ (uranium). Furthermore, the dependence of the results on the size of the target materials and the muon exposure time was considered.
 
 \begin{figure}[H]
\begin{subfigure}{.47\textwidth}
  \centering
  \includegraphics[width=1.0\linewidth]{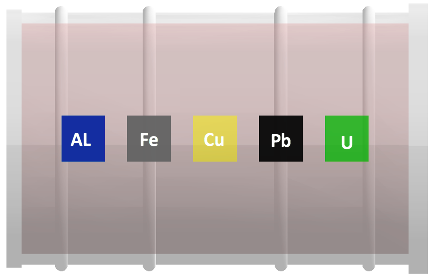}  \caption{}
  \label{fig:sub-first4}
\end{subfigure}
\begin{subfigure}{.51\textwidth}
  \centering
  \includegraphics[width=1.0\linewidth]{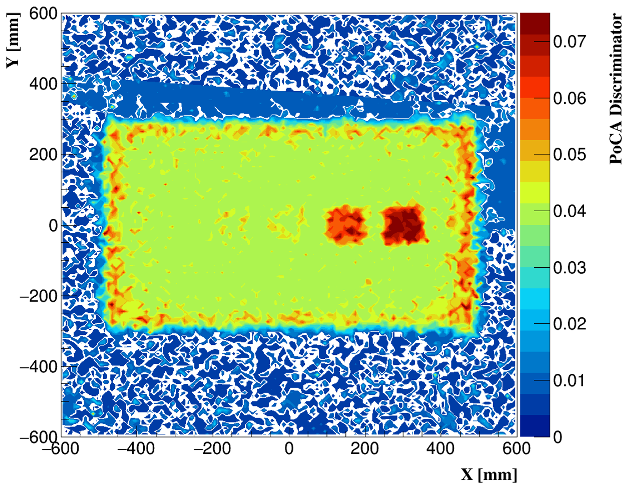}  
  \caption{}
  \label{fig:sub-second4}
\end{subfigure}

\begin{subfigure}{.51\textwidth}
  \centering
  \includegraphics[width=1.0\linewidth]{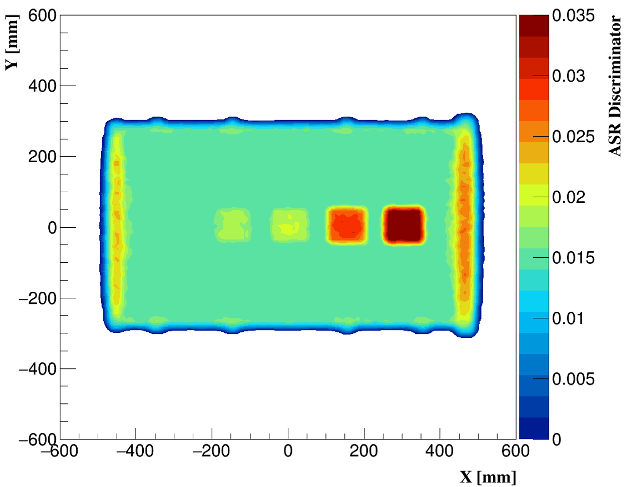}  
  \caption{}
  \label{fig:sub-third4}
\end{subfigure}
\begin{subfigure}{.51\textwidth}
  \centering
  \includegraphics[width=1.0\linewidth]{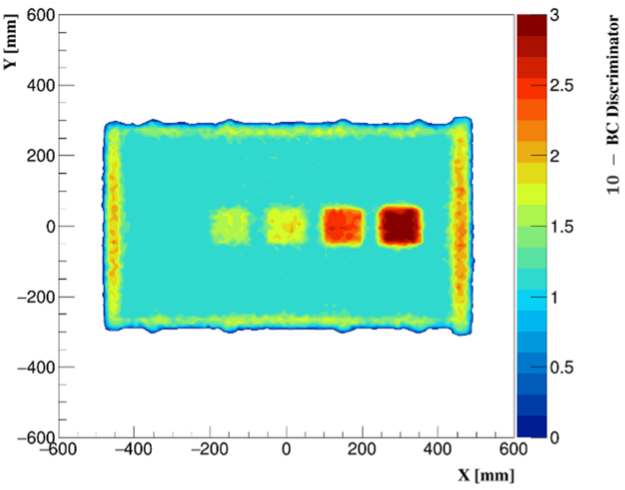}  
  \caption{}
  \label{fig:sub-forth4}
\end{subfigure}
\caption{(a) Target materials inside the simulated drum. X-Y slice outputs through the 3D density map from applying the (b) PoCA, (c) ASR, and (d) BC algorithms respectively. The exposure time was 30 days equivalent. The BC algorithm considered the 38 most scattered tracks per voxel (N).}
\label{figure8_label}
\end{figure}

\subsubsection{Size and location dependence}
 In order to avoid the limitations of some reconstruction algorithms in cases of short exposure time, this section will compare the performance of all methods with fixed muon exposure time. The performance of each method is represented here by the CNR value between two different regions after 30 days of muon exposure time. The sliced outputs shown in Figure~\ref{figure8_label} are taken through the 3D density maps along the centre of the drum.
 In this case the target materials are of size 10 $\times$ 10 $\times$ 10 cm$^3$ inside the waste drum.
 The outputs clearly show that all three reconstructon methods are able to locate high-Z materials (U and Pb) shielded by the concrete matrix, whereas the PoCA method failed to identify the medium-Z target materials (Cu and Fe). 

For the low-Z target material, all methods are unable to separate the aluminium cube from the background, which is expected because the aluminium has an almost similar density to the concrete.

\begin{figure}[ht]
\begin{subfigure}{.5\textwidth}
  \includegraphics[width=1.01\linewidth]{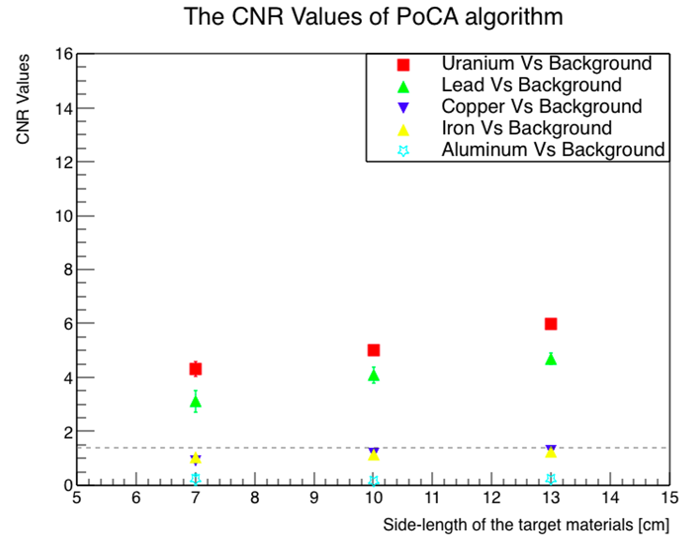}  \caption{}
  \label{fig:sub-first5}
\end{subfigure}
\begin{subfigure}{.5\textwidth}
  \includegraphics[width=1.01\linewidth]{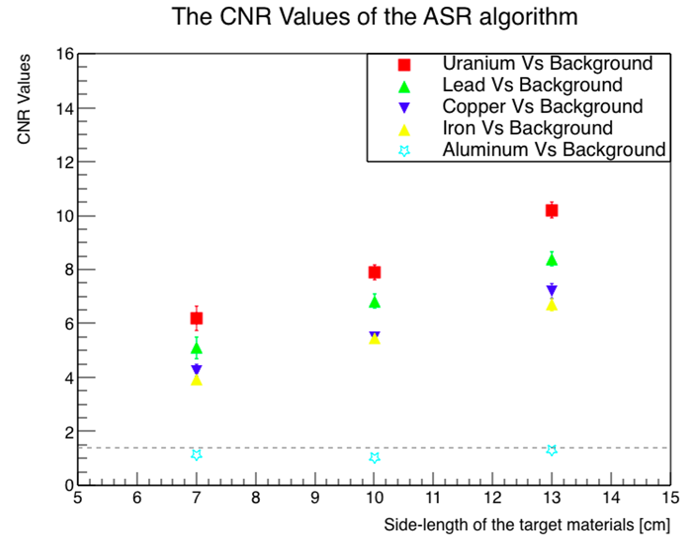}  
  \caption{}
  \label{fig:sub-second5}
\end{subfigure}
  \begin{center}
      \begin{subfigure}{.5\textwidth}
  \includegraphics[width=1.01\linewidth]{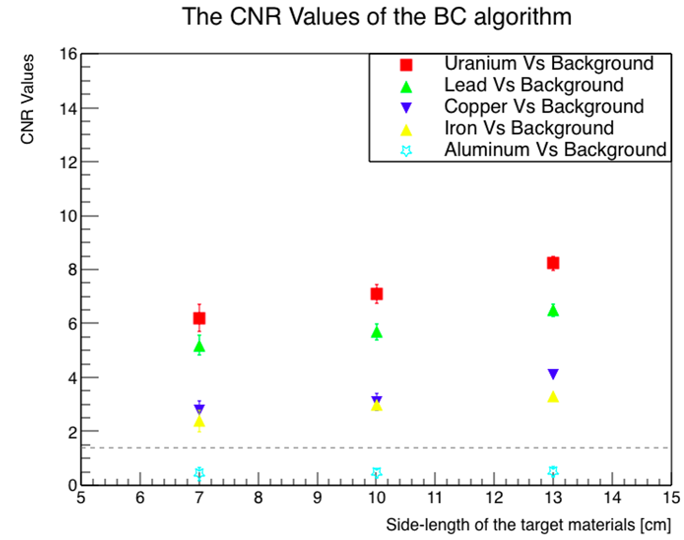}  \caption{}
  \label{fig:sub-third5}
\end{subfigure}
  \end{center}
  \caption{Comparison of the CNR values of the (a) PoCA, (b) ASR and (c) BC algorithms when differentiating between different target materials and background for target materials with side lengths of 7, 10, and 13~cm. Results are  for 30 days of muon exposure time. The vertical dashed line represents the minimum CNR value used to distinguish the target material inside the drum.}
  \label{figure9_label}
\end{figure}
Figure \ref{figure9_label} shows the CNR results of all algorithms used after 30 days of muon exposure when comparing the five target materials, namely uranium, lead, copper, iron, and aluminium individually against the regions that have background signals. The BC and ASR algorithms demonstrate very similar performance when comparing the regions that contained a high-Z material (uranium) cube against the background regions.
In the case of a 10~cm cube the BC method produces a slightly lower CNR value of 7.1~$\pm$~0.34 compared to the CNR value of 7.9~$\pm$~0.25 produced by the ASR algorithm. 

The PoCA method is affected by the single-scattering assumption, which leads to adding more noise which reduces the CNR values between the compared regions inside the drum. For example, for 10~cm sided cubes the PoCA algorithm has only been able to distinguish the uranium and lead from the background with relatively low CNR values of 5.0~$\pm$~0.2 and 4.1~$\pm$~0.3, respectively.
\newline The results from comparing the regions containing copper and uranium reveal that the ASR algorithm is the most capable of differentiating between medium-Z and high-Z materials with a CNR value of 5.35~$\pm~$0.1, which is approximately 34$\%$ better than the CNR value produced for the comparable regions by the BC method. In terms of the size dependence, the CNR results from comparing the target regions showed that the ASR is more likely to be affected by the target region's size. For example, comparing lead cubes against background regions showed almost a 65~$\%$ increase in the CNR values from 5.1~$\pm~$0.4 to 8.4~$\pm~$0.26 when the side-length of the lead cube increased from 7~cm to 13~cm. 
\subsubsection{Dependence on muon exposure time}
\begin{figure}[hpt!]
\begin{subfigure}{.5\textwidth}
  \centering
  \includegraphics[width=1.0\linewidth]{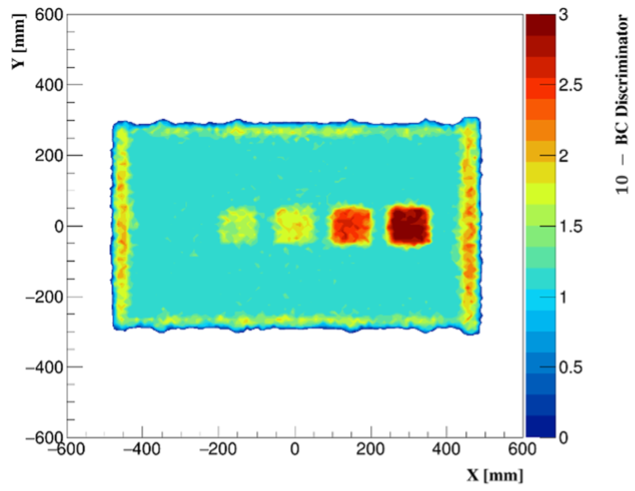}  \caption{}
  \label{fig:sub-first44}
\end{subfigure}
\begin{subfigure}{.5\textwidth}
  \centering
  \includegraphics[width=1.0\linewidth]{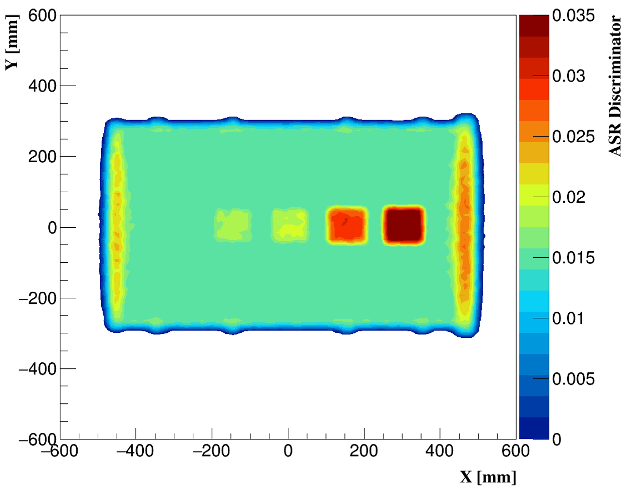}  
  \caption{}
  \label{fig:sub-second44}
\end{subfigure}

\begin{subfigure}{.5\textwidth}
  \centering
  \includegraphics[width=1.0\linewidth]{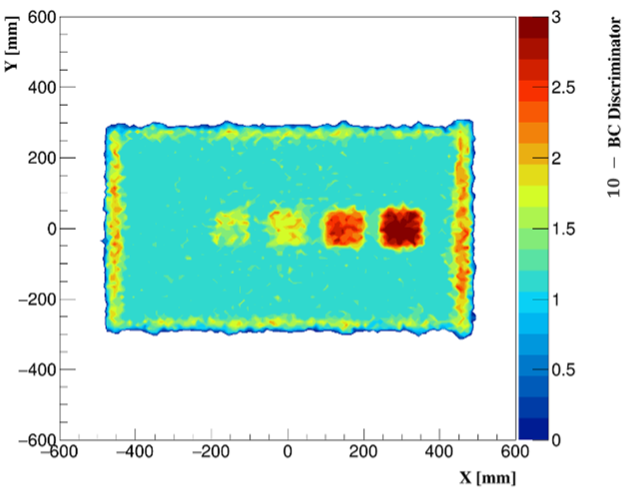}  
  \caption{}
  \label{fig:sub-third44}
\end{subfigure}
\begin{subfigure}{.5\textwidth}
  \centering
  \includegraphics[width=1.0\linewidth]{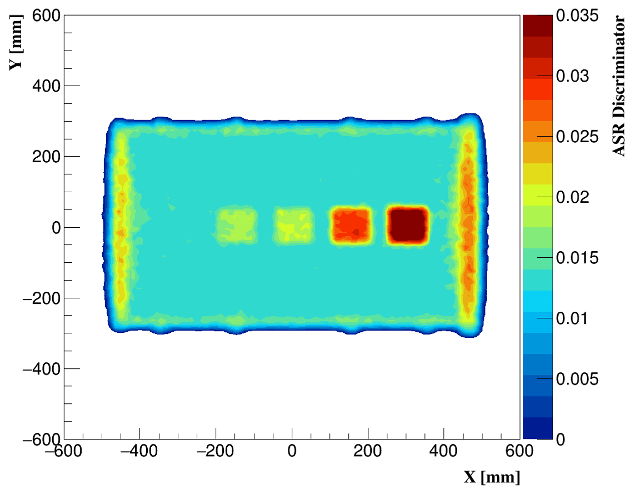}  
  \caption{}
  \label{fig:sub-fourth44}
\end{subfigure}
\begin{subfigure}{.5\textwidth}
  \centering
  \includegraphics[width=1.0\linewidth]{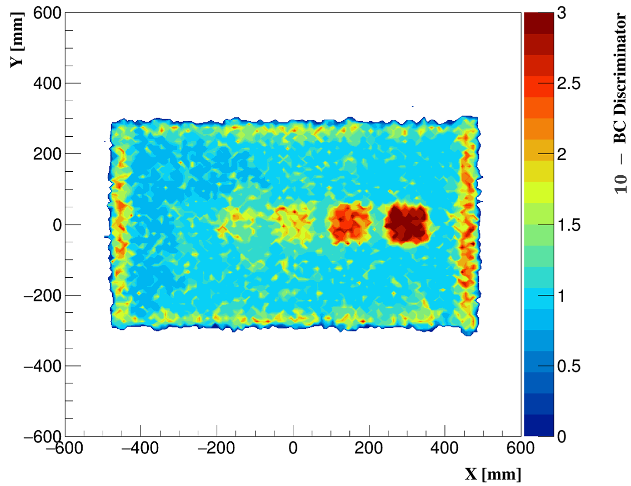}  
  \caption{}
  \label{fig:sub-fifth44}
\end{subfigure}
\begin{subfigure}{.5\textwidth}
  \centering
  \includegraphics[width=1.0\linewidth]{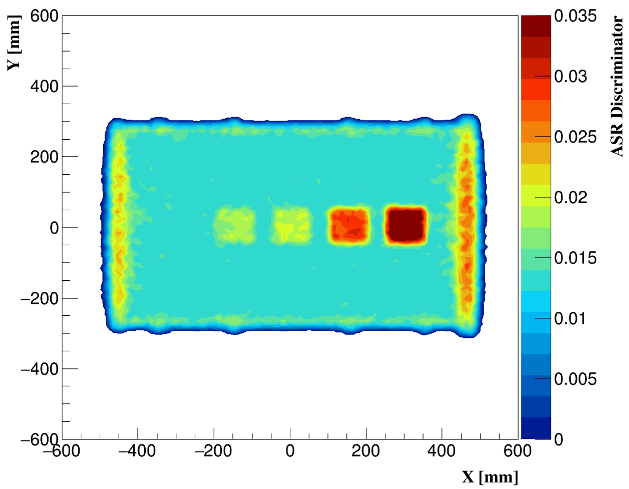}  
  \caption{}
\end{subfigure}
\caption{Comparison of feature resolution of the target materials with a side length of 10 cm reconstructed by the BC algorithm (left) and by the ASR algorithm (right) after 16 (top), 8 (middle), and 4 (bottom) days of muon exposure. The BC algorithm considered the  24, 12 and 6 most scattered tracks per voxel for 16, 8 and 4 days of exposure time, respectively.}
\label{figure10_label}
\end{figure}

An additional variable of muon exposure time, i.e. the number of muons contributing to each simulation, must be considered to
to fully interpret the CNR values mentioned above. 
Figure~\ref{figure10_label} shows the output density maps produced by the ASR and BC algorithms after 4, 8 and 16 days of exposure time for the same target materials inside the small waste drum. 


\begin{figure}[hbt!]
\begin{subfigure}{.5\textwidth}
  \includegraphics[width=1.1\linewidth]{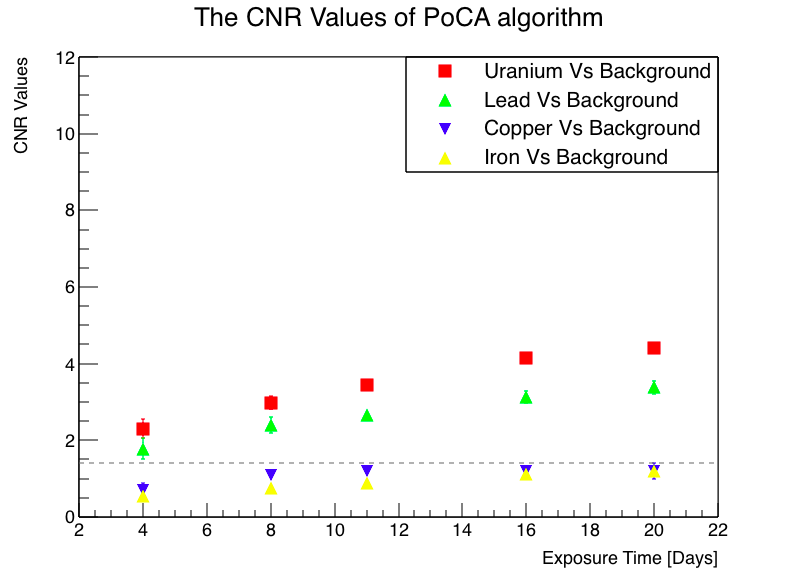}  \caption{}
  \label{fig:sub-first55}
\end{subfigure}
\begin{subfigure}{.5\textwidth}
  \includegraphics[width=1.02\linewidth]{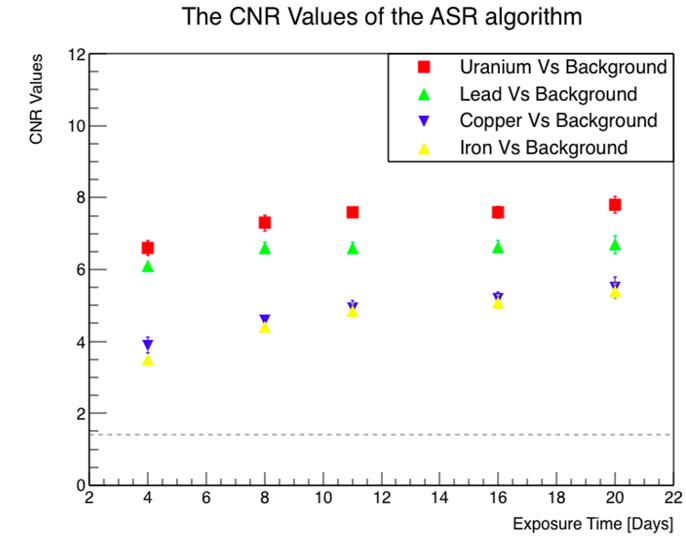}  
  \caption{}
  \label{fig:sub-second55}
\end{subfigure}
  \begin{center}
      \begin{subfigure}{.5\textwidth}
  \includegraphics[width=1.02\linewidth]{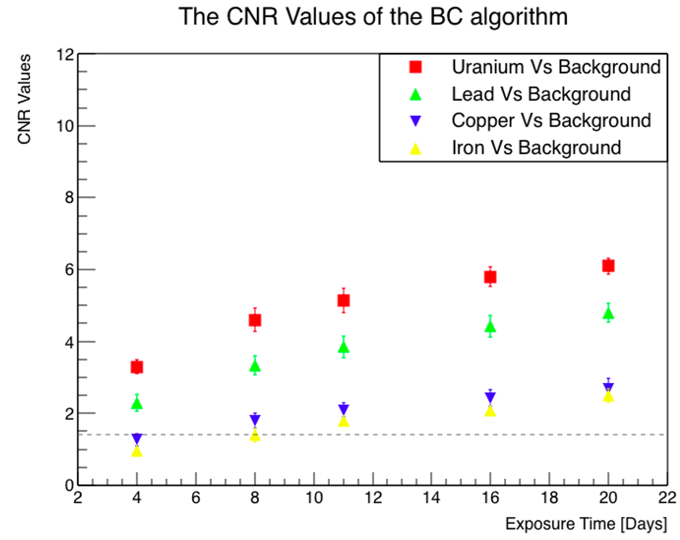}  \caption{}
  \label{fig:sub-third55}
\end{subfigure}
  \end{center}
  \caption{Comparison of the CNR values produced by the (a) PoCA, (b) ASR and (c) BC algorithms for different materials of 10~~cm side-length as a function of the muon exposure time. The vertical dashed line represents the minimum CNR value used to distinguish the target material inside the drum. }
\label{figure11_label}
\end{figure}
The ASR algorithm produces images of the target materials with greater clarity even when the exposure time is as short as 4 days. This clarity in the reconstructed image is ascribed to the fact that outsider events with large scattering angles are not included in the weights of the voxel. The ASR method maintained its capability of distinguishing between the uranium cube and the background region with CNR values of 7.3~$\pm$~0.2 after 8 days of muon exposure time.

The performance of the BC algorithm is significantly reduced when the drum’s content image is reconstructed in 4 days’ time (see Figure~\ref{figure10_label}(e)). For instance, the medium-Z materials (Cu and Fe) have become indistinguishable from the background regions, with CNR values of just 1.3~$\pm$~0.2 and 0.97~$\pm$~0.12 for the copper and iron targets, respectively.  However, high-Z materials still can be differentiated from the background with CNR values of 3.3$~\pm$~0.2 and 2.3~$\pm$~0.23 when comparing uranium and lead respectively against the background regions. A summary of the CNR values for 10~cm side-length cubes of uranium, lead, copper and iron as a function of muon exposure time is illustrated in Figure~\ref{figure11_label}.

Generating tomographic images of the target materials can be achieved with fewer cosmic muons by using the ASR algorithm. By using the ASR method, the MST system can separate uranium and lead from background regions in only six hours of muon exposure time with CNR values of 3.1$~\pm$~0.2 and 2.5$~\pm$~0.2 respectively.
\newpage
\subsubsection{Influence of the spatial resolution}
In addition to muon momentum information, the discriminator values in each voxel of all algorithms described in section \ref{section3} are calculated by mainly using the information of the muons' scattering angles. The BC and PoCA algorithms use ($\theta$) and the ASR algorithm uses ($\theta_x$ and $\theta_y$). Firstly, the CNR values between the target regions listed in tables \ref{table:1} and \ref{table:2} were measured according to the primary scattering angle information obtained from the simulated default system setting described in section \ref{subsection4.1}. The angular resolution depends on the hit resolution of the detectors. To understand the influence on the angular resolution of the hit position resolution, a study was performed degrading the drift chamber and the RPC hit position resolutions from 2 mm to 4 mm and 350 micron to 700 micron respectively. 

\begin{figure}[h]
\centering
\includegraphics[width=.487\textwidth]{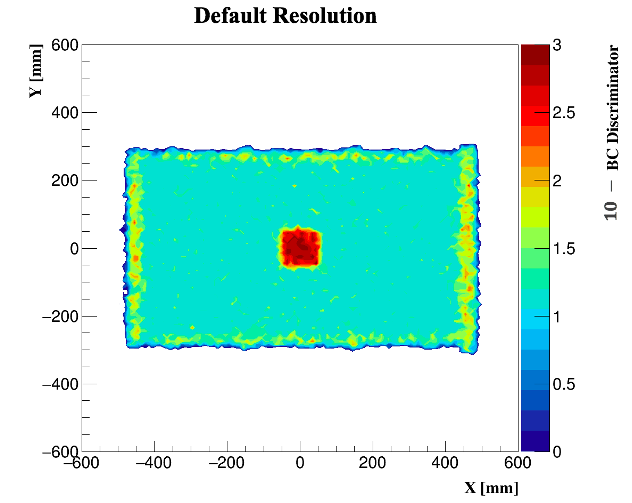}\quad
\medskip
\includegraphics[width=.47\textwidth]{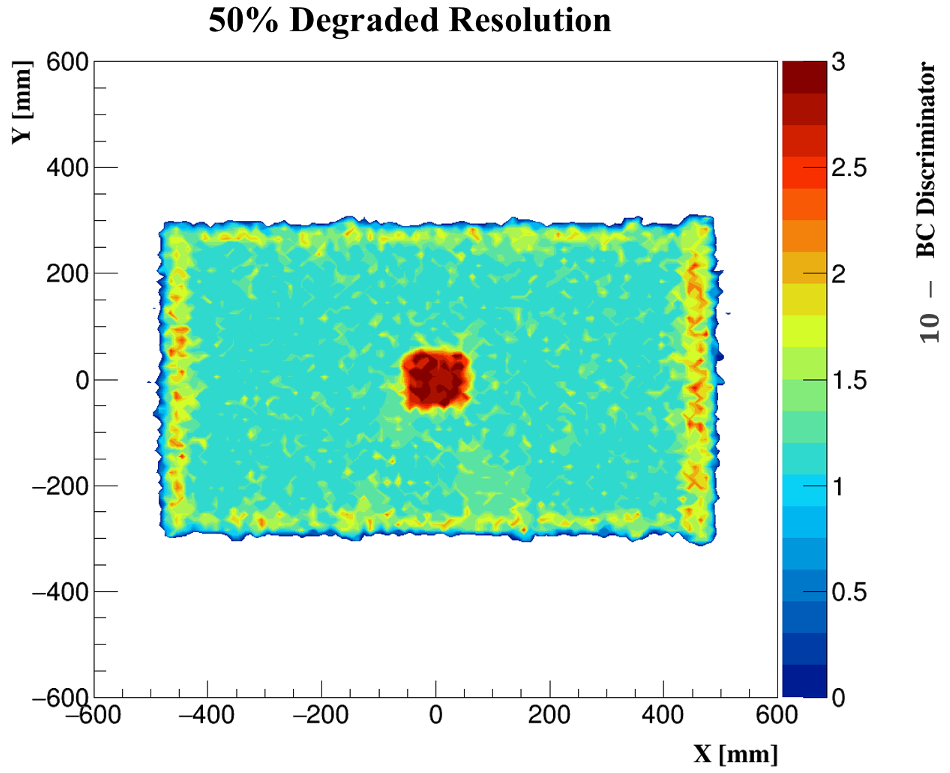}\quad

\caption{A 10 cm side length tungsten cube placed in the centre of a simulated cement matrix nuclear waste drum. The 2D projected BC outputs of the same cube inside the small drum using (left) the default detector system and (right) the detector system after degrading the RPC and DC position resolutions to 0.525 and 3 mm respectively. The voxels containing the background regions have less variation of the reconstructed image by the default system. The exposure time was eight days equivalent.}
\label{Smeared_label}
\end{figure}

Figure \ref{Smeared_label} shows examples of the BC outputs of a 10 cm side-length tungsten cube before and after degrading the spatial resolutions by 50$\%$, i.e. the RPC's resolution to 0.525~mm and the DC's resolution to 3~mm, respectively. The cube structure is easily distinguishable in both figures. However, the reconstructed image of the cube appears to be sharper when using the default detectors' resolutions. Furthermore, there is a slight variation in the discriminator values of the background matrix in the degraded resolution image of the drum.

\begin{figure}[hbt!]
\begin{subfigure}{.52\textwidth}
  \includegraphics[width=1.02\linewidth]{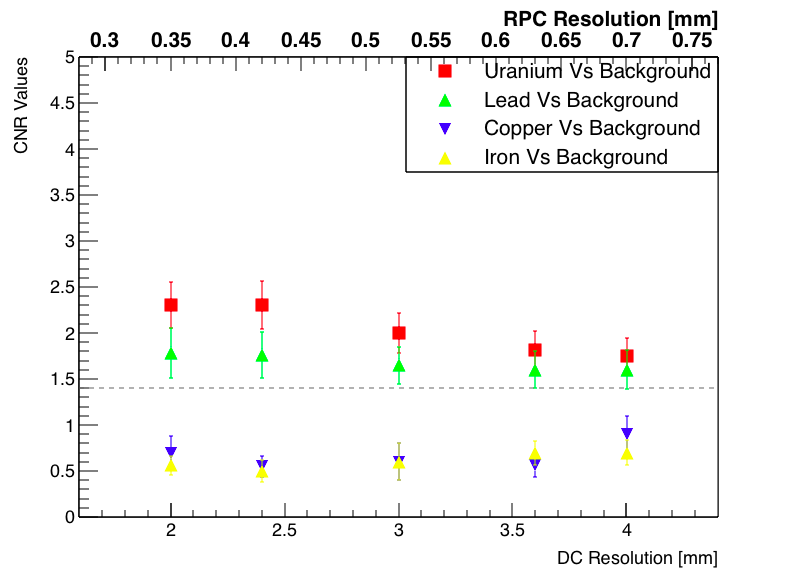}  \caption{}
  \label{fig:sub-first555}
\end{subfigure}
\begin{subfigure}{.52\textwidth}
  \includegraphics[width=1.02\linewidth]{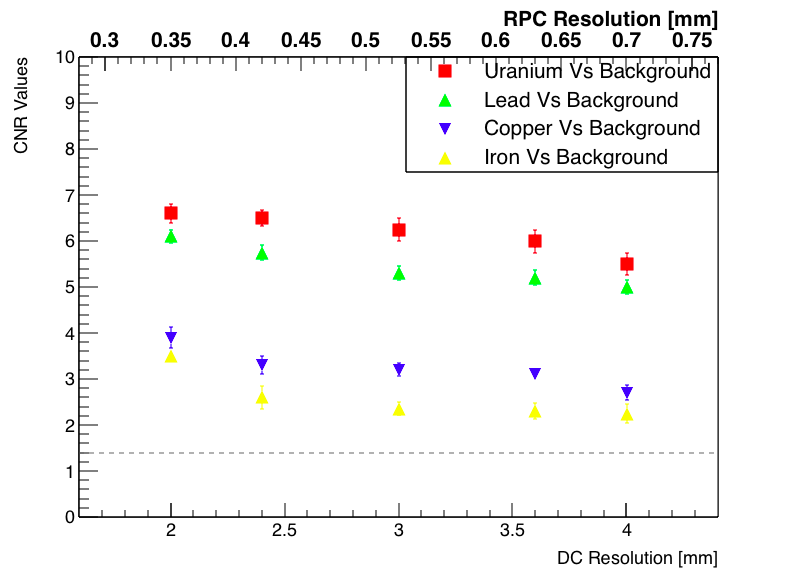}  
  \caption{}
  \label{fig:sub-second555}
\end{subfigure}
  \begin{center}
      \begin{subfigure}{.52\textwidth}
  \includegraphics[width=1.02\linewidth]{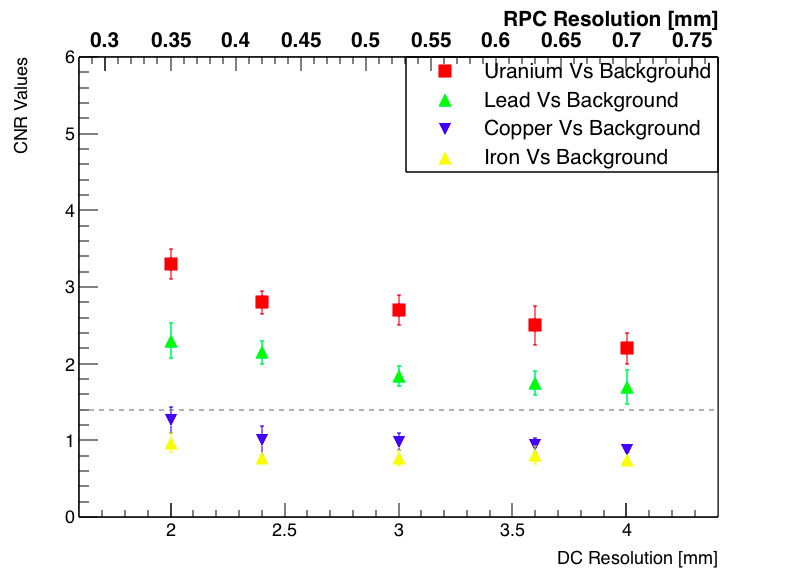}  \caption{}
  \label{fig:sub-third555}
\end{subfigure}
  \end{center}
  \caption{Comparison of the CNR values produced by the (a) PoCA, (b) ASR and (c) BC algorithms for different materials of 10~~cm side-length as a function of the muon detector resolutions. The vertical dashed line represents the minimum CNR value used to distinguish the target material inside the drum. The exposure time was 4 days equivalent. }
\label{figure111_label}
\end{figure}

The results of additional simulations of the MST detector with degraded spatial resolutions have been performed with both fixed muon exposure time (4 days) and material size ( 10 cm side-length). Figure \ref{figure111_label} shows the CNR results of comparing the 4 target materials against the background regions, using the PoCA in (a), ASR in (b), and BC in (c).

The quality of the reconstructed images has been slightly affected by degrading the detectors' resolutions, especially when comparing the medium-Z. materials against the background. The ability of the detector system to differentiate between lead with 10 cm side-length and equally-sized background regions is slightly degraded from a CNR value of 6.1$~\pm$~0.14 to 5$~\pm$~0.15, when using the ASR algorithm. However, even with poor resolutions of the detector, all the target materials reconstructed by the ASR can be distinguished, e.g., the copper and the iron cubes can be separated from the background regions with CNR values of 2.7$~\pm$~0.16 and 2.25$~\pm$~0.2, respectively. 

Regardless of muon exposure time, the detector's capability of reconstructing images of the target materials using the PoCA and BC methods has been kept above the mCNR value for the high-Z materials against the background regions. The detector's system has not significantly lost its detection capability even when using RPCs with a resolution of 700 micron and DCs with a resolution of 4 mm. Hence, our detector setup is able to generate tomographic images of shielded high-Z and medium-Z materials with fewer muons when using the ASR method ( 4 days of muon exposure time).

\begin{figure}[h]
\centering
\includegraphics[width=.74\textwidth]{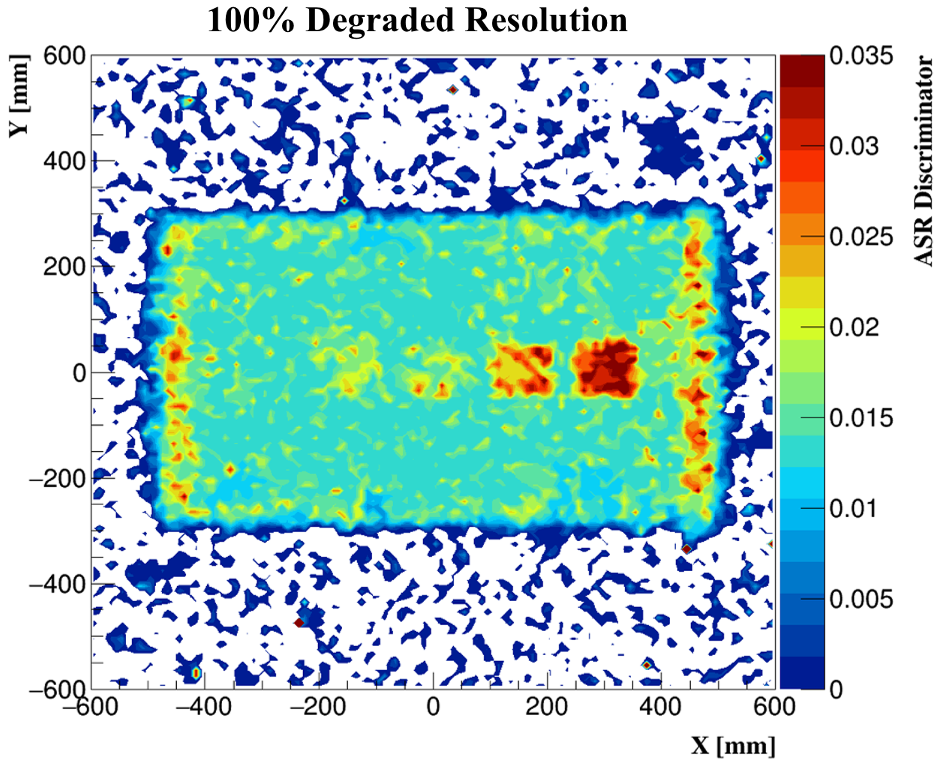}\quad

\caption{ The 2D projected output of the target materials with a side length of 10 cm reconstructed by the ASR algorithm using the detector system after degrading the RPC and DC position resolutions to 0.7 and 4 mm respectively. The exposure time was six hours equivalent.}
\label{SixHours_label}
\end{figure}

With less exposure time ( 24 hours), the detector system starts to lose its capability in separating copper and iron from the background with a CNR value of 1.35~$\pm$~0.15 and 1.0~$\pm$~0.23, when using the ASR method. However, the ability of the system to discriminate high-Z materials, such as uranium, from the background can still be achieved in only six hours of exposure time with a CNR value of 2.4~$\pm$~0.25 (see Figure \ref{SixHours_label}).
\newpage
\subsection{Applying the CNR test to a large V/52 CASTOR drum}
This section presents the results of an extended study of the feature resolutions of the reconstructed images in the presence of thicker and denser shields. The feature resolution is quantitatively represented here by the CNR value between a single target basket that is half-loaded with UO$_2$, copper, and lead pellets and the eight fully-loaded UO$_2$ baskets that surround the investigated basket.
\begin{figure}[hpt]
\begin{subfigure}{.4\textwidth}
  \includegraphics[width=1.03\linewidth]{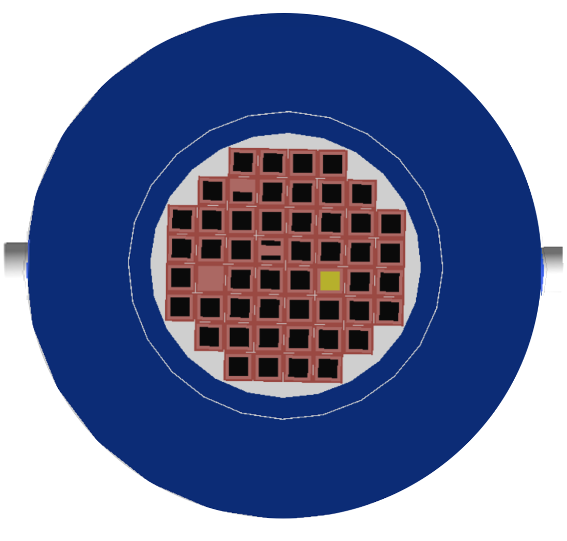}  \caption{}
  \label{fig:sub-firstCAS}
\end{subfigure}
\begin{subfigure}{.5\textwidth}
  \includegraphics[width=1.0\linewidth]{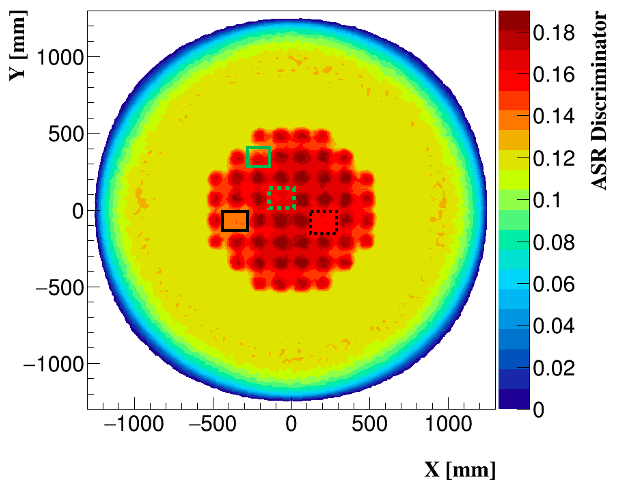} 
  \caption{}
  \label{fig:sub-secondCAS}
\end{subfigure}
      \begin{subfigure}{.5\textwidth}
  \includegraphics[width=1.0\linewidth]{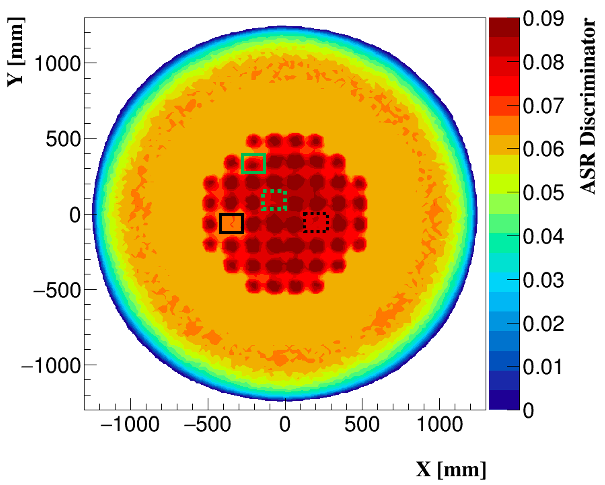}  \caption{}
  \label{fig:sub-thirdCAS}
\end{subfigure}
      \begin{subfigure}{.5\textwidth}
  \includegraphics[width=1.0\linewidth]{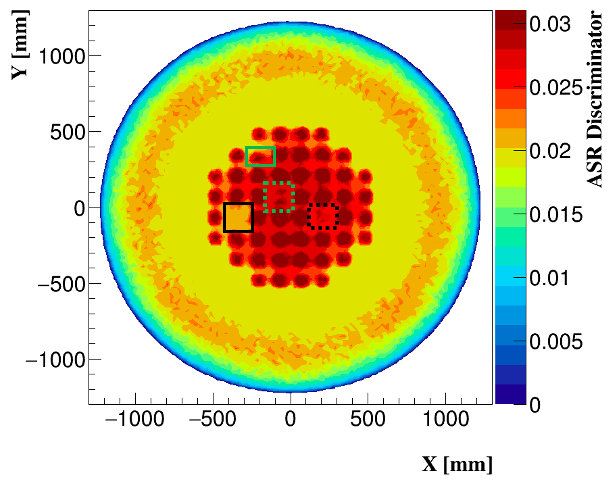}  \caption{}
  \label{fig:sub-forthCAS}
\end{subfigure}
  \caption{(a) top-view of the V/52 CASTOR showing four baskets contain irregularity in their contents. Comparison of the feature resolutions of the cask's contents produced by the ASR algorithm when considering (b) 75$\%$, (c) 50$\%$ and (d) 25$\%$ of the ASR discriminator in each voxel, respectively. The solid and dashed green boxes indicate the half-loaded baskets, while the solid and dashed black boxes indicate the baskets that contain no pellets and copper pellets, respectively.The exposure time was 30 days equivalent.}
  \label{figure12_label}
\end{figure}

The PoCA and BC algorithms could not separate the empty basket from the eight neighbouring, fully loaded baskets with CNR values of 0.9~$\pm$~0.13 and 0.3~$\pm$~0.1, respectively. This is likely caused by the PoCA assumption of approximation of multiple scattering in one single scatter, which results in poor approximation of the scattering locations. This also impacts on the feature resolution capabilities of the BC algorithm which uses the scattering location of PoCA algorithm. This assumption mainly affects the vertical positions of the muon scattering, as the momentum of cosmic muons is much larger in the vertical direction; hence, fluctuations in the clustering values occur.

The ASR algorithm succeeds in minimising the smearing noise that results from the PoCA assumption. Additional quantiles of 25$\%$ and 50$\%$ of the ASR discriminator distributions in each voxel are considered (see Figure~\ref{figure12_label}). Comparing the empty basket with the eight surrounding fully loaded baskets 
produces a CNR value of 2.8~$\pm$~0.25 when 75$\%$ of the discriminator distributions were taken in each voxel. However, considering the 25$\%$ quantile of each voxel distribution improves the CNR value about 78$\%$ to 5.0~$\pm$~0.3. Hence, hereon all quoted CNR results will use the ASR[25$\%$] value.
\begin{figure}[hbt]
\begin{center}
      \includegraphics[scale=0.35]{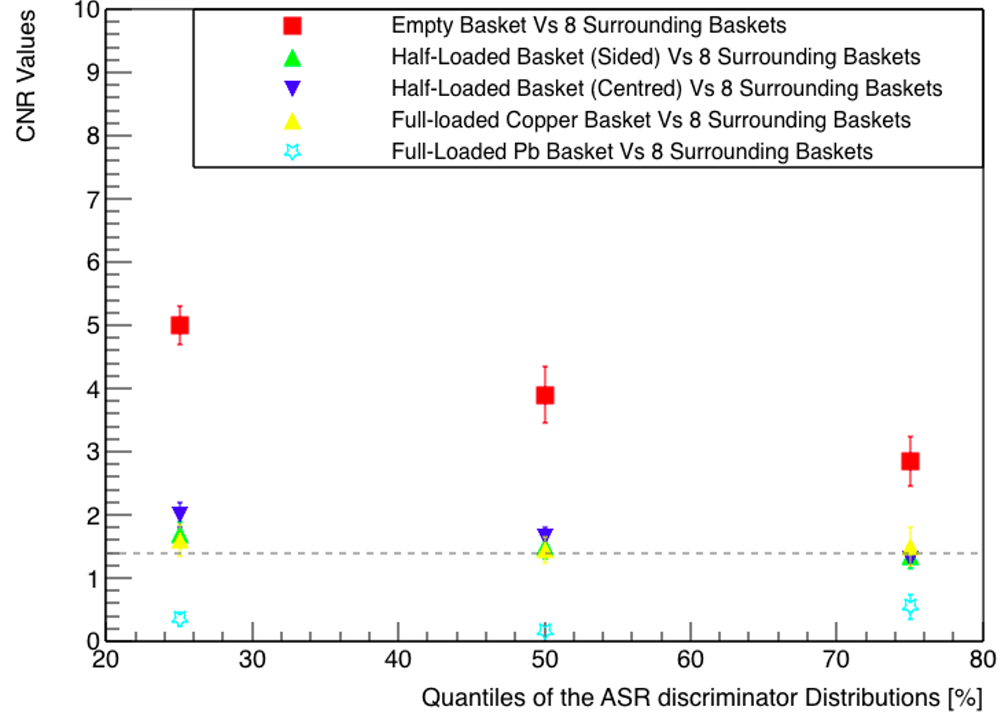}
     \caption{Comparison of the CNR values produced by the ASR discriminator when considering 25$\%$, 50$\%$ and 75$\%$ of voxel's distributions after 30 days of muon exposure. The vertical dashed-line represents the minimum CNR value in which the comparable regions could be distinguished from each other.} 
     \label{figure14_label}
 \end{center} 
\end{figure}

\begin{figure}[hpt]
\begin{subfigure}{.48\textwidth}
  \centering
  \includegraphics[width=1.0\linewidth]{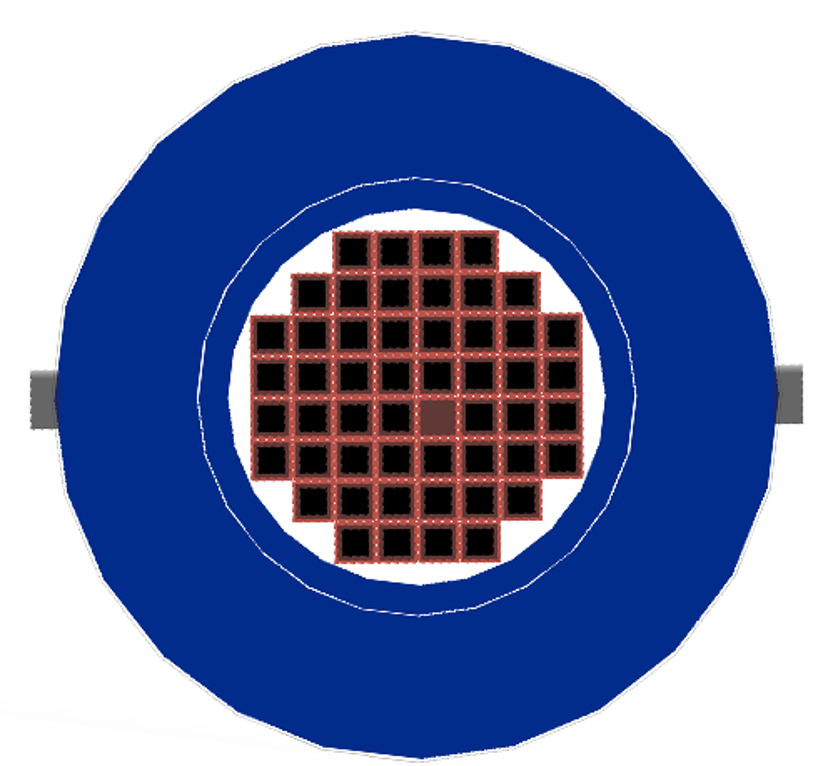}  \caption{}
  \label{fig:sub-CAS0}
\end{subfigure}
\begin{subfigure}{.5\textwidth}
  \centering
  \includegraphics[width=1.0\linewidth]{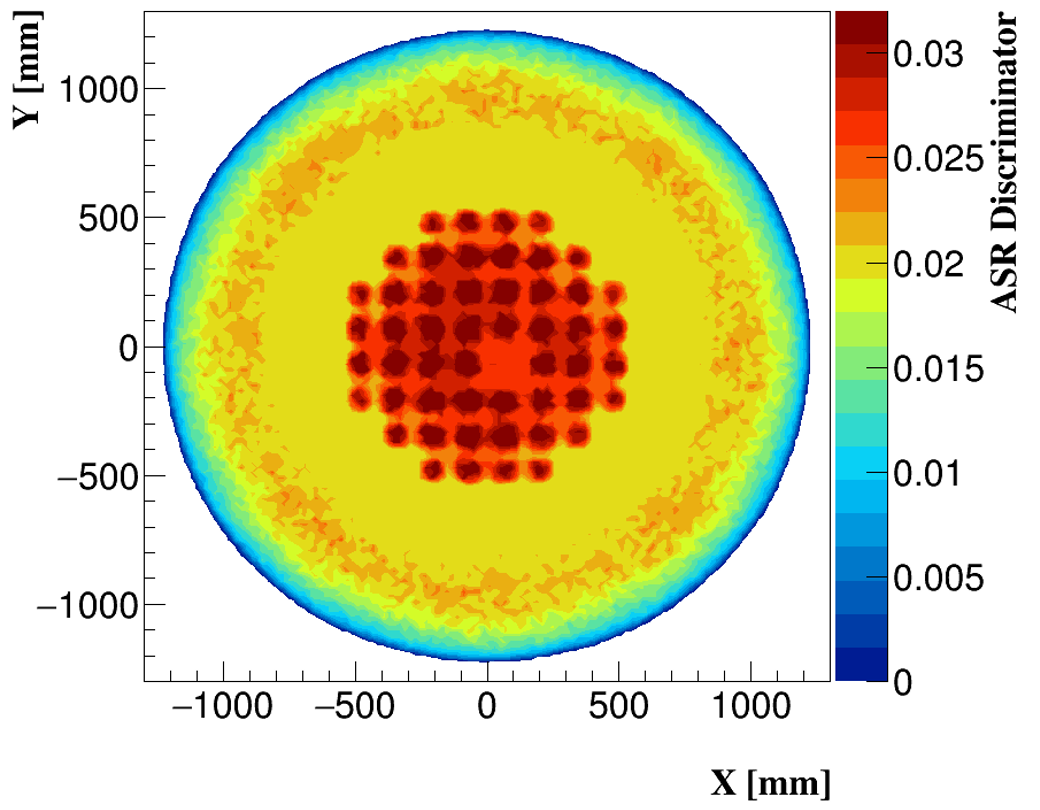}  \caption{}
  \label{fig:sub-CAS1}
\end{subfigure}
\begin{subfigure}{.5\textwidth}
  \centering
  \includegraphics[width=1.0\linewidth]{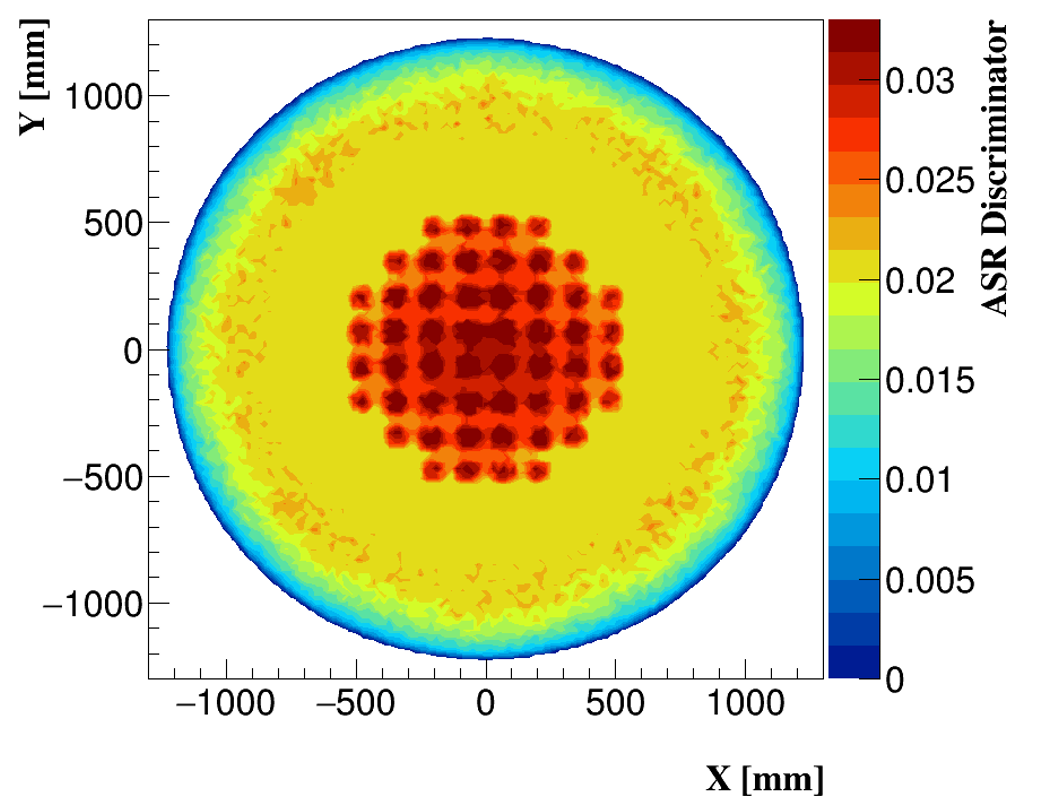}
  \caption{}
  \label{fig:sub-CAS2}
\end{subfigure}
\begin{subfigure}{.5\textwidth}
  \centering
  \includegraphics[width=1.0\linewidth]{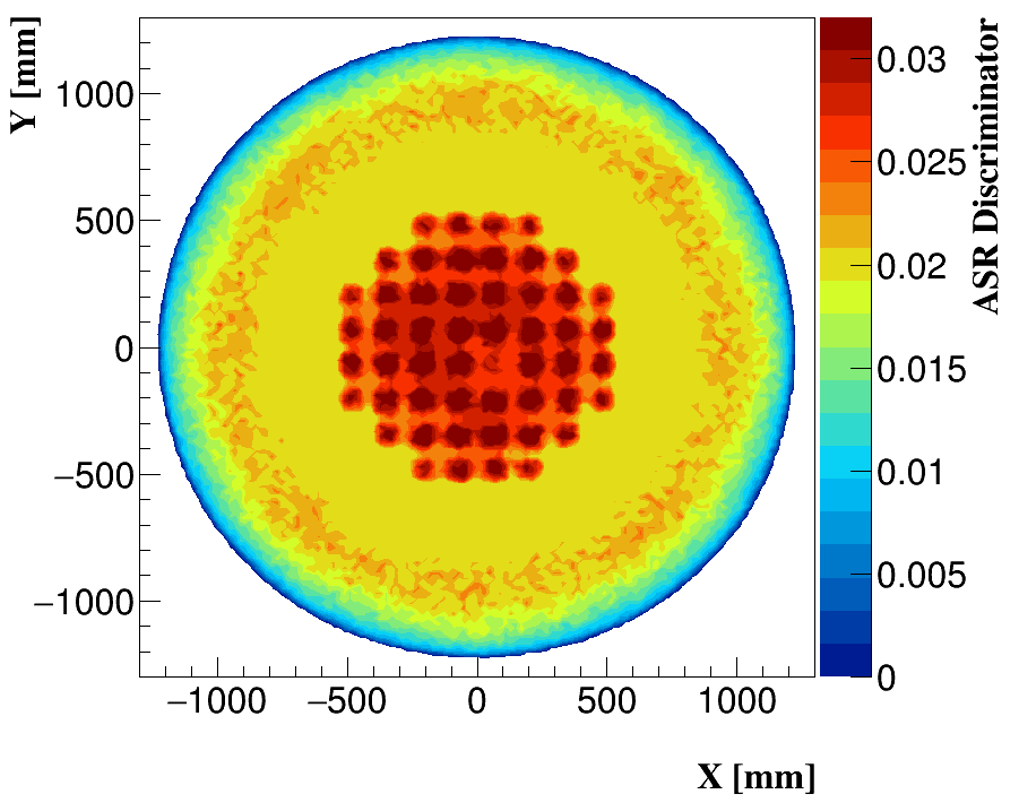} 
  \caption{}
  \label{fig:sub-CAS3}
\end{subfigure}
\begin{subfigure}{.5\textwidth}
  \centering
  \includegraphics[width=1.0\linewidth]{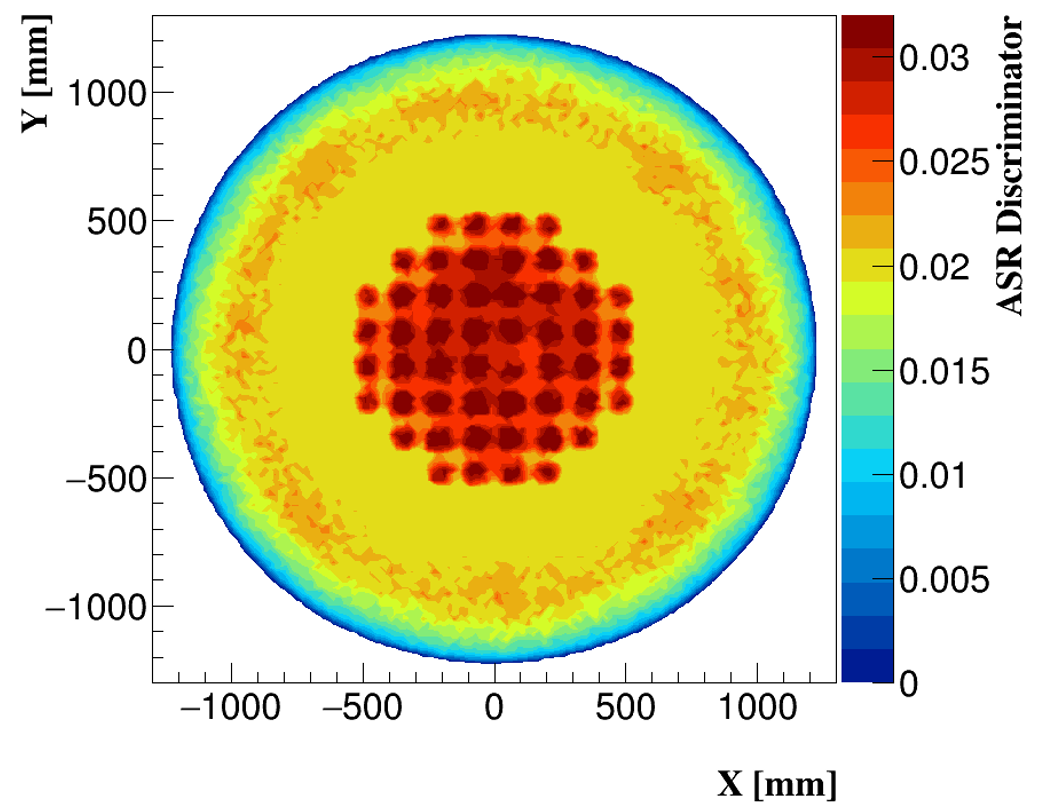} 
  \caption{}
  \label{fig:sub-CAS4}
\end{subfigure}
\begin{subfigure}{.5\textwidth}
  \centering
  \includegraphics[width=1.0\linewidth]{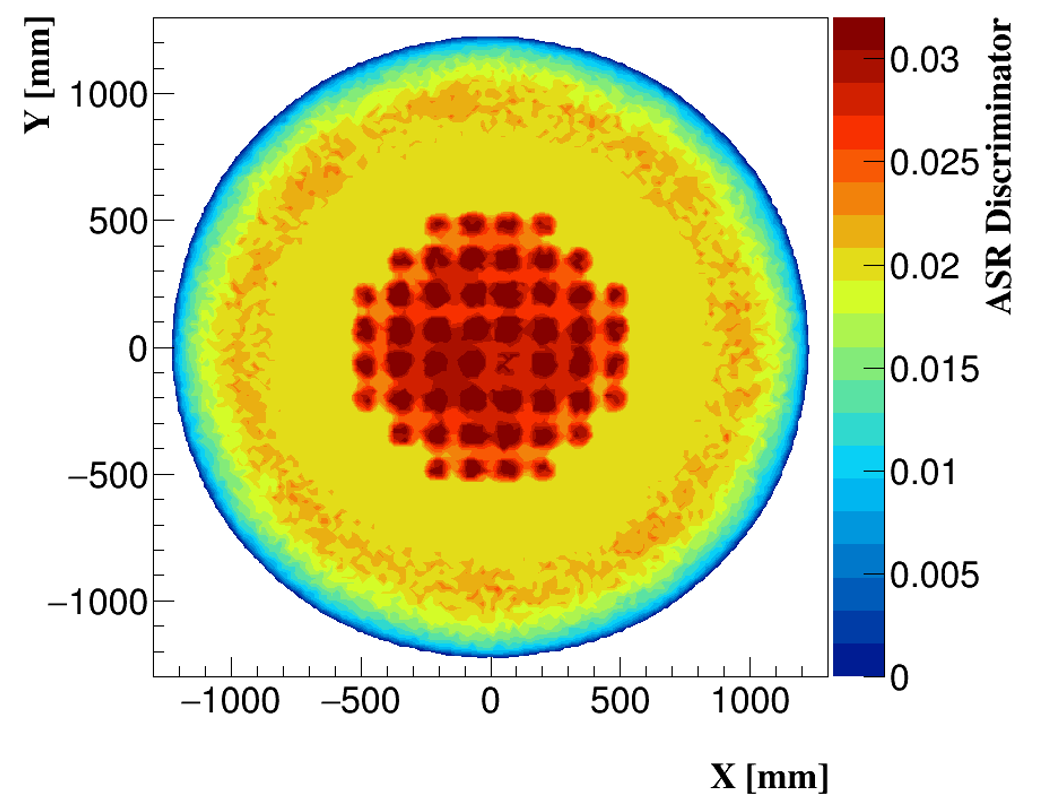} 
  \caption{}
  \label{fig:sub-CAS5}
\end{subfigure}
\caption{Comparison of the feature resolutions of the contents of basket number 30 when only 25$\%$ of the ASR discriminator’s distributions are considered in each voxel. All figures indicate basket number 30 accommodating (b) no pellets, (c) fully-loaded with Pb pellets, (d) fully-loaded with Cu pellets, (e) half-sided and (f) half-centred. The exposure time was 30 days equivalent.}
\label{figure13_label}
\end{figure}
Figure~\ref{figure14_label} shows a comparison using 25$\%$, 50$\%$ and 75$\%$ quantiles of the ASR distributions inside each voxel after 30 days of muon exposure time. 
As expected, the CNR values increase as the quantile decreases, the improvement being greatest for those material combinations where there is the greatest level of discrimination.

The X-Y projections of the 52 baskets accommodating 51 full-loaded baskets with UO$_2$ fuel assemblies and one basket (number 30) accommodating the target materials are shown in Figure~\ref{figure13_label}.
Testing the size reliance of the target material was achieved by comparing half-loaded baskets to the eight baskets surrounding it. This shows the ability of the ASR method to separate the irregular contents of the basket. The CNR values from comparing half-unloaded (centered) and half-unloaded (sided) baskets to the eight fully loaded baskets appear to be just above the minimum distinguishable CNR level of 1.9~$\pm$~0.2 and 1.6~$\pm$~0.3, respectively.  As expected, the regions of the basket filled with lead pellets and the surrounding baskets are not distinguishable (CNR = 0.35~$\pm~0.05$) due to the similarity of lead and UO$_2$ densities.

\begin{figure}[h]
\begin{subfigure}{.5\textwidth}
  \includegraphics[width=1.0\linewidth]{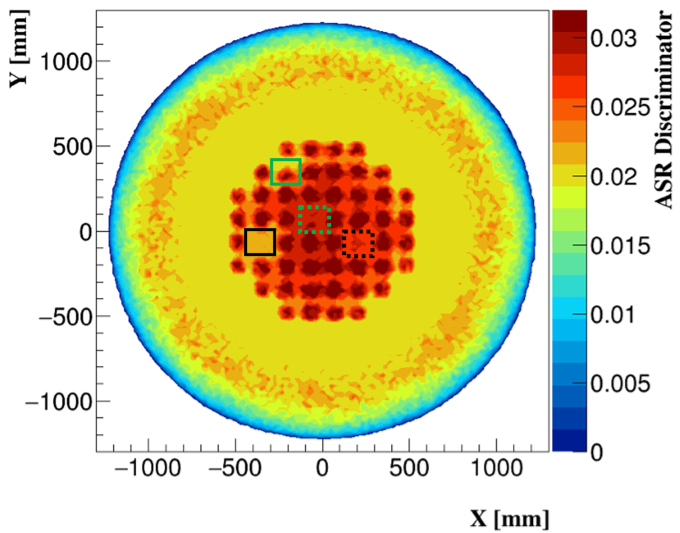}  \caption{}
  \label{fig:sub-firstCAS11}
\end{subfigure}
\begin{subfigure}{.5\textwidth}
  \includegraphics[width=1.0\linewidth]{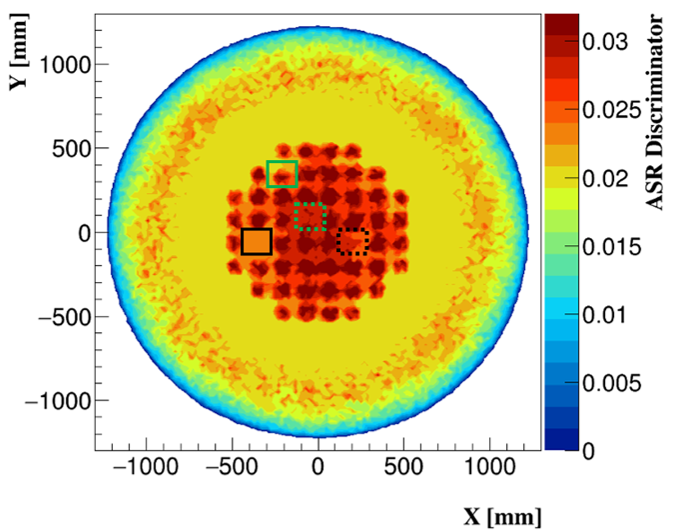}  
  \caption{}
  \label{fig:sub-secondCAS22}
\end{subfigure}
      \begin{subfigure}{.5\textwidth}
  \includegraphics[width=1.0\linewidth]{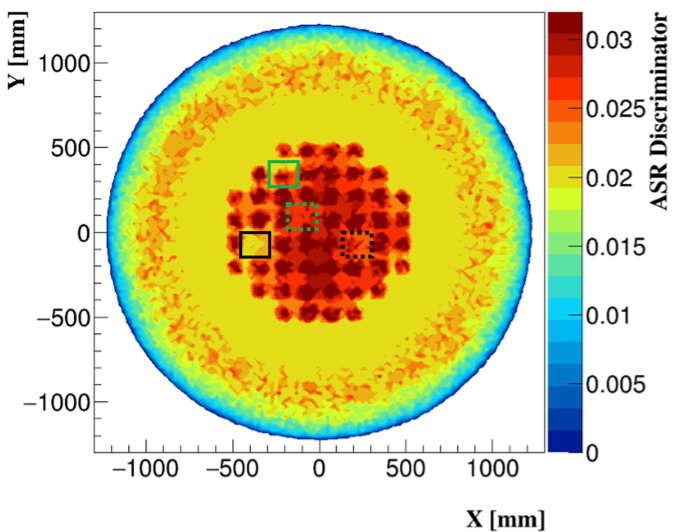}  \caption{}
  \label{fig:sub-thirdCAS33}
\end{subfigure}
      \begin{subfigure}{.5\textwidth}
  \includegraphics[width=1.0\linewidth]{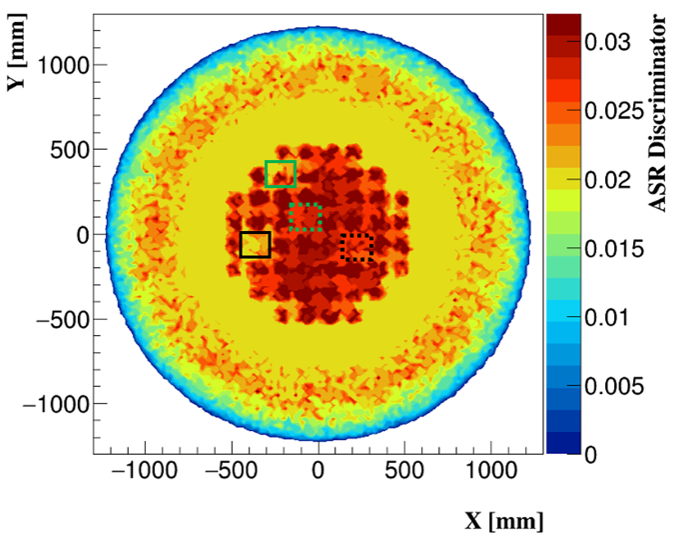}  \caption{}
  \label{fig:sub-forthCAS44}
\end{subfigure}
  \caption{Comparison of the imaging the cask's contents produced via the ASR algorithm when considering 25$\%$ of the ASR discriminator in each voxel after (a) 20, (b) 11, (c) 8 and (d) 4 days of muon exposure time, respectively. The solid and dashed green boxes in indicate the half-loaded baskets, while the solid and dashed black boxes indicate the baskets that contain no pellets and copper pellets, respectively.}
  \label{figure15_label}
\end{figure}

It is possible to use the CNR test to rapidly evaluate how the detector performance might be affected when one of the operating conditions, such as the muon exposure time, is changed. The output density map of the ASR discriminator is shown in Figure~\ref{figure15_label} for between 4 and 20 days of muon exposure time. After 11 days of muon exposure time, the noise in the reconstructed density map renders the half-loaded baskets indistinguishable, with CNR values of 1.05~$\pm$~0.2 and 0.93~$\pm$~0.25 when half of the fuel assemblies are unloaded from the centre of the basket and from the side of the basket, respectively. 

After 4 days of muon exposure time, the density maps are constrained by the detector’s angular resolution and the fuel assemblies inside the image are smeared by the neighbouring baskets. Finally, in terms of the detector resolutions, the MST detector kept its performance in locating the empty basket from the surrounding neighboring fully-loaded baskets, after degrading the system's resolutions. For instance, after sixteen days of muon exposure time, the ability of the detector system to separate the empty basket from the eight surrounded baskets is degraded from a CNR value of 2.8~$\pm$~0.2 to 2.2~$\pm$~0.115, 2.0~$\pm$~0.23, and 1.65~$\pm$~0.25, when degrading the detectors' resolutions by factors of 20, 50 and 80$\%$, respectively.
\section{Conclusion}
We conducted a quantitative method to evaluate the performance of our MST detector system in differentiating between high-Z and medium-Z materials. The CNR method was applied to assay three reconstruction algorithms in terms of their ability and limitations to differentiate between chosen target materials of different sizes, positioned in different locations inside a small-cemented matrix drum and a large-scale CASTOR V/52 cask.

For the small drum, the CNR results conclude that the BC and the ASR algorithms were sufficiently capable of locating and differentiating between regions containing high-and medium-Z materials with side-lengths of 7, 10, and 13~cm against 4 regions containing background signal. The BC method performance grows gradually as the target material size increases with CNR values of 6.2~$\pm~$0.5, 7.1~$\pm$~0.34 and 8.23~$\pm$~0.26 for 7, 10, and 13 cm side length uranium cubes, respectively. However, this performance degrades significantly, and the ability of the BC method is constrained, when the muon exposure time was shortened to four days with CNR values of 1.3$\pm$0.2 and 0.97$\pm$0.12 for copper and iron compared to background regions. This rendered the medium-Z materials indistinguishable from the background.

The ASR algorithm was shown to be more efficient in investigating target materials in a short time. All the materials investigated by the ASR lie above the distinguishable level (except Aluminium) even when the muon exposure time is as short as 4 days, with CNR values of  6.6~$\pm$~0.2, 6.1~$\pm$~0.14, 3.9~$\pm$~0.22 and 3.5~$\pm$~0.12 when comparing 10 cm side-length of uranium, lead, copper, and iron against background regions. Despite degrading the detector resolution to 700 micron for the RPCs and 4 mm for the DCs, the detector system has maintained the ability to separate the copper cube with 10 cm side-length from the background regions with a CNR value of 2.7~$\pm$~0.16 after 4 days of muon exposure.

For the larger V/52 cask, the PoCA and the BC method failed to locate any of the irregular baskets that had 100$\%$ and 50$\%$ of their capacity unloaded, with CNR values of 0.9~$\pm$~0.13 and 0.3~$\pm$~0.1. The ASR method has been shown to be a good candidate for examining large and well-shielded materials. The ASR discriminator worked well to decrease the effects of the PoCA single-scatter assumption, and it demonstrated the ability to locate any irregularity within the fuel assemblies, such as empty, half-unloaded, and basket composite copper pellets. Our detector system can locate any empty basket, whether it is located in the side or the centre of the cask, with a CNR value of 5.0~$\pm$~0.3. Despite degrading the detector resolution to 525 micron for the RPCs and 3 mm for the DCs, the detector system has displayed the capability of the system in identifying the missing contents of any basket with a CNR value of 2.0~$\pm$~0.23.
The ability of the ASR algorithm to detect half-unloaded baskets is limited by the angular resolution of the detector when the cask’s content is investigated with an exposure time of 11 days or less.
\acknowledgments
This project has received funding from the Euratom research and training programme 2014-2018 under grant agreement No 755371.
\bibliographystyle{JHEP}
\bibliography{RefJinstDec.bib}

\end{document}